\documentclass[trsc]{informs4}
\OneAndAHalfSpacedXI

\usepackage{natbib}
\bibpunct[, ]{(}{)}{,}{a}{}{,}%
\def\bibfont{\small}%

\TheoremsNumberedThrough
\EquationsNumberedThrough

\usepackage{amsmath,amssymb,amsfonts}
\usepackage{mathrsfs}
\usepackage{mathtools}
\usepackage{multirow}
\usepackage{longtable}
\usepackage{subfig}
\usepackage{graphicx} 
\usepackage{enumerate}
\usepackage{makecell}
\usepackage[bookmarks=true,linkcolor=red,citecolor=blue,urlcolor=green,colorlinks=true, breaklinks]{hyperref}
\usepackage{booktabs}
\usepackage{algorithm}
\usepackage{algorithmic}
\usepackage{threeparttable}
\usepackage{color}
\usepackage{bbm}
\usepackage{lscape}
\usepackage{rsfso}
\usepackage{dsfont}
\usepackage{mathpazo}
\usepackage{graphics, diagbox, caption}

\usepackage{verbatim} 	
\usepackage{lineno}

\renewcommand{\theARTICLETOP}{}
\newcommand{\prob}{FMTP-BD}

\newcommand{\mcA}{\mathcal{A}}
\newcommand{\mcV}{\mathcal{V}}
\newcommand{\mcS}{\mathcal{S}}
\newcommand{\mcB}{\mathcal{B}}
\newcommand{\mcE}{\mathcal{E}}
\newcommand{\mcC}{\mathcal{C}}
\newcommand{\mcN}{\mathcal{N}}

\newcommand{\mcT}{\mathcal{T}}

\newcommand{\mcD}{\mathcal{D}}
\newcommand{\mcU}{\mathcal{U}}

\newcommand{\mcR}{\mathcal{R}}
\begin{document}
\RUNAUTHOR{}
\RUNTITLE{The FMTP-BD}
\TITLE{The Freight Multimodal Transport Problem with Buses and Drones: An Integrated Approach for Last-Mile Delivery}
\ARTICLEAUTHORS{%
 \AUTHOR{E Su}
 \AFF{School of Management, Huazhong University of Science and Technology, Wuhan 430074, China,\\
 \EMAIL{es1996@hust.edu.cn}}
 \AUTHOR{Hu Qin}
 \AFF{School of Management, Huazhong University of Science and Technology, Wuhan 430074, China,
 \EMAIL{tigerqin1980@qq.com}}
  \AUTHOR{Jiliu Li\footnote{Corresponding author}}
 \AFF{School of Management, Northwestern Polytechnical University, Xi’an 710072, China, \EMAIL{jiliuli@nwpu.edu.cn}}
 \AUTHOR{Rui Zhang}
 \AFF{Leeds School of Business, University of Colorado Boulder, Colorado 80309, USA, \EMAIL{rui.zhang@colorado.edu}}
}





\ABSTRACT{

\textit{\textbf{Problem definition:}} This paper proposes a novel freight multimodal transport problem with buses and drones, where buses are responsible for transporting parcels to lockers at bus stops for storage, while drones are used to deliver each parcel from the locker to the corresponding customer. The integrated bus-drone system synergistically expands drone service coverage using the bus network to ensure efficient final delivery. Minimizing the total operational costs while satisfying customer demands necessitates the joint optimization of parcel assignments and drone flights. \textit{\textbf{Methodology/results:}} We model the problem into a compact mixed-integer linear programming formulation and propose an integer programming formulation with exponentially many variables. To address real-world scale instances, we propose a Branch-Price-and-Benders-Cut algorithm for this non-deterministic polynomial-time (NP)-hard problem. This algorithm, integrating column generation and Benders decomposition within a Branch-and-Bound framework, is developed to obtain optimal or near-optimal solutions. Additionally, we introduce algorithmic enhancements aimed at accelerating the convergence of the algorithm. Computational experiments on instances generated from real-world bus data demonstrate that the proposed algorithms outperform CPLEX regarding both efficiency and solution quality. Moreover, our approaches can lead to over 6\% cost savings compared to situations where we determine parcel assignments and drone flights sequentially. \textit{\textbf{Implications:}} We evaluate the environmental advantages of integrating buses and drones, study the impact of different cost parameters in the system, and investigate the impact of the parcel locker configuration on performance. These findings provide valuable managerial insights for urban logistics managers, highlighting the potential of the integrated bus-drone system to improve traditional last-mile delivery.

}




\maketitle
\section{Introduction}
The swift growth of online commerce has introduced significant challenges to the courier industry, especially regarding the management of the surging number of parcel deliveries. With the global parcel volume surpassing 161 billion shipments in 2022, as reported by the Global Parcel Shipping Index \citep{Dies2022}, courier companies are under pressure to find efficient and reliable delivery solutions. Despite this, traditional logistics, primarily reliant on trucking, face limitations such as traffic congestion, logistical complexities, and inadequate parking spaces \citep{Sun2023}.
 
To overcome these challenges, urban multimodal transport has emerged as a promising solution to enhance freight mobility. This approach advocates for the sharing of public transportation for both passengers and freight \citep{Ghilas2018, Kizil2023}. Leveraging the excess capacity of buses during off-peak hours for parcel transportation has shown potential as an efficient and cost-effective option to meet the growing demand for deliveries. By utilizing the existing bus network, parcels can be consolidated at bus terminals and transported using buses to parcel lockers at the bus stops for storage. Incorporating buses expands delivery coverage, leading to cost savings and efficiency gains, while dedicated bus lanes help alleviate road congestion and reduce emissions. Examples of such initiatives can be found in Binzhou, China \citep{Yang2023}, and La Rochelle, France \citep{Masson2017}. Nevertheless, the use of buses for parcel transportation presents certain limitations in last-mile delivery. This is mainly because buses follow fixed lines and schedules that may not always align with the specific destination of individual parcels. As a result, traditional methods continue to dominate due to the flexibility of home delivery offered by company-owned fleets, even though establishing and maintaining a dedicated fleet may pose significant cost considerations.


To achieve lower economic and environmental costs, drone operation has been proposed as a new solution for last-mile delivery in urban logistics \citep{He2022, Cheng2024}. Amazon Prime Air \citep{amazonair} is one of the most famous attempts to utilize drones for this purpose. In 2021 \citep{chamber}, the Dallas-Fort Worth Metroplex became the first major U.S. metropolitan area where a commercial drone delivery service is approved and available, representing a significant step forward for the drone delivery industry. Nowadays, \citet{Walmart} is offering their drone delivery for up to 75\% of the Dallas-Fort Worth population (over 6.6 million). Drones offer the advantage of direct delivery, bypassing road conditions and ensuring quick access to customers, which improves efficiency compared to ground transportation. This approach also brings additional benefits, such as delivering packages to balconies even when customers are absent, thus enhancing the customer experience and reducing failed deliveries \citep{Pugliese2021}. However, drone delivery has limited service coverage due to the limited flight range and carrying capacity.

\begin{figure}[tb]
\vspace{-.4cm}
	\centering
	\includegraphics[width = 0.7\textwidth]{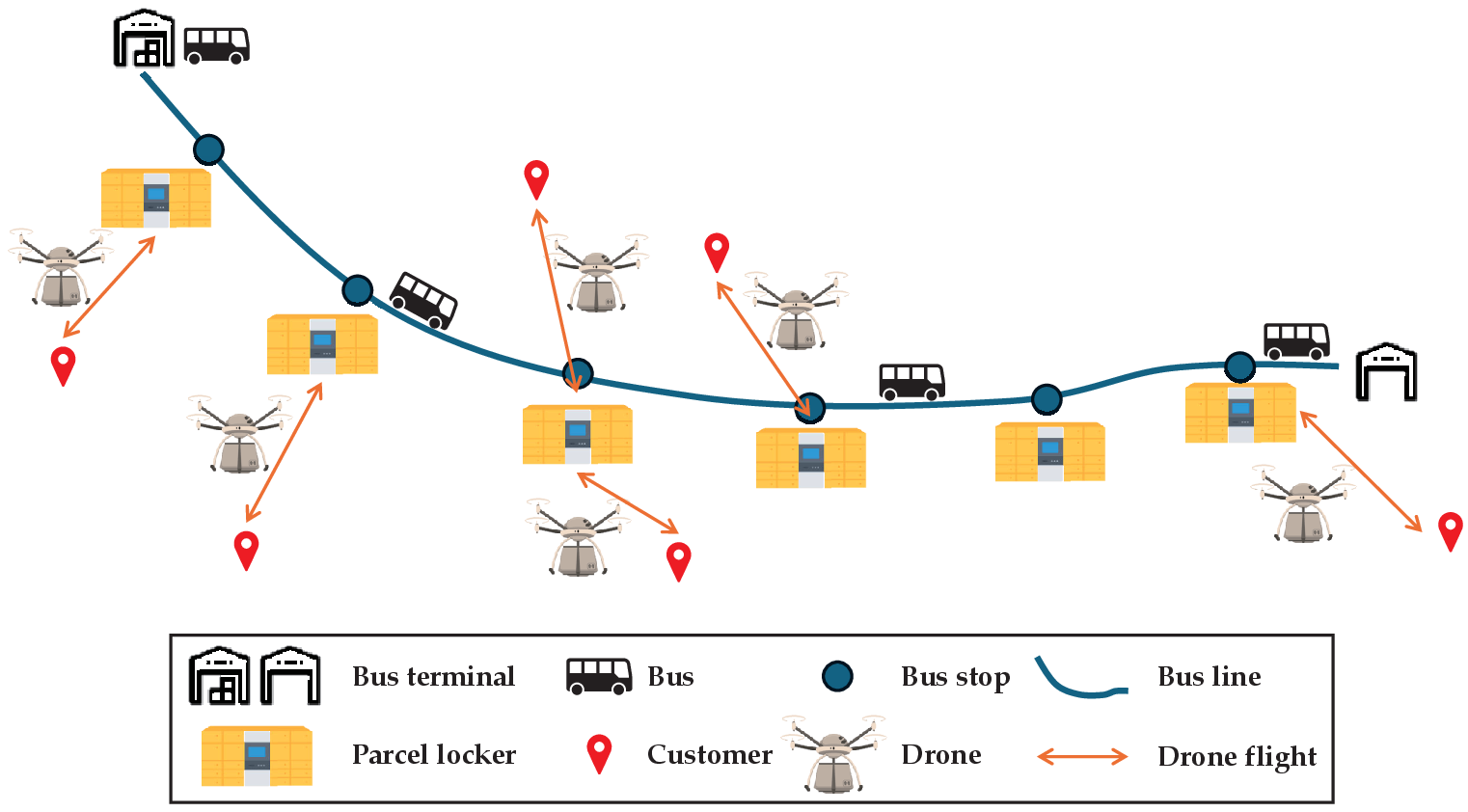}
	\caption{Illustration of the Integrated Bus-Drone System.}	\label{figIntro}	
 \vspace{-.5cm}
\end{figure}

Rather than pairing drones with trucks, which is the most common way in the literature \citep{Murray2015, Wang2019, Poikonen2019}, we are interested in a freight multimodal transport system that integrates buses and drones. Integrating drones with buses not only extends the limited service coverage of drones but also addresses the aforementioned issues caused by the misalignment between fixed bus lines and parcel destinations. To this end, this paper introduces an integrated system that combines the bus network and drones operated by logistics companies. Parcels are consolidated at the bus terminal, transported through buses to parcel lockers at bus stops, and then delivered to customers by drones, as illustrated in Figure~\ref{figIntro}. While we might examine each component in the system separately, a comprehensive way is needed to account for the interaction among the included components. Moreover, bus and drone operations should be jointly optimized for the system's efficiency and cost-effectiveness. Our model and solution approaches provide valuable tools to advance the current understanding of freight multimodal transport systems for last-mile delivery.


The main contributions of this paper are as follows.
\begin{itemize}
	\item This paper proposes the freight multimodal transport problem with buses and drones (\prob). The integrated system combines buses and drones to expand service coverage for last-mile delivery, creating a synergistic optimization of urban logistics.
	\item We present a compact mixed-integer linear programming (MILP) formulation aimed at minimizing the total operation costs of the system while fully satisfying customer demands. To tackle larger instances, we propose an integer programming (IP) formulation with two sets of exponentially many variables: one set corresponds to parcel assignments, and the other one corresponds to drone flights. Note that each set of variables is a set of columns in a mathematical programming formulation. We refer to this IP formulation as the two-column-based formulation.
 
	\item Built upon the two-column-based formulation, this paper develops solution approaches that combine column generation and Benders decomposition within the branch-and-bound framework, employing efficient labeling algorithms to generate columns and intelligent branching strategies to obtain integer solutions. We also propose algorithmic enhancements to improve the tractability.  Computational experiments are conducted on instances using real-world bus data. In terms of computational efficiency and solution quality, the results not only demonstrate the superiority of the proposed approaches but also validate the effectiveness of the algorithmic enhancements. The proposed algorithms significantly outperform CPLEX. Moreover, our approaches can lead to over 6\% cost savings compared to the situations where we determine parcel assignments and drone flights sequentially. 
 
 \item Sensitivity analyses on various scenarios offer valuable managerial insights and highlight the potential of the proposed bus-drone integrated system. First, we evaluate the environmental advantages of integrating buses and drones.  Key findings reveal that the bus-drone system
leads to significantly lower carbon emissions, highlighting the emerging benefits and operational
efficiency of the integrated system. Next, we study the impact of different cost parameters in the system. Our results show total costs are more sensitive to bus operation costs than drone operation costs, underscoring the need for efficient parcel management. It also highlights the significance of enhancing the efficiency of drone delivery, especially under high demand, to manage costs effectively. Lastly, we investigate the impact of the parcel locker configuration on the performance. Our findings show that increasing the number of lockers initially reduces the total cost, but the cost reduction plateaus once a certain density is reached, beyond which additional lockers offer diminishing returns unless the number of customers increases. 
\end{itemize}

The remainder of this paper is organized as follows. We review the relevant literature in Section \ref{secReview}. Section \ref{secProblem} formally defines the \prob\ and formulates it as a compact MILP model. We introduce the solution method in Section \ref{secAlgorithm}, followed by the algorithmic enhancements in Section \ref{secImprovement}. The computational results and managerial insights are reported in Section \ref{secExperiments}. Section \ref{secConclusion} concludes the work and discusses future research. All mathematical proofs are provided in \ref{appProve}.

\section{Literature Review}	\label{secReview}
This section reviews the literature and focuses on two core areas that are central to this paper: public transport-based logistics and drone delivery problems.

\subsection{Public Transport-Based Logistics}

Public transport-based logistics has emerged as an innovative solution for efficient and sustainable urban freight transport by leveraging shared public transport infrastructure for both passengers and freight. These systems offer significant economic and environmental benefits, which we explore at both the strategic and operational levels in the academic research that follows.

\cite{Masson2017} explored the integration of passenger and freight services on public transport networks. They proposed an adaptive large neighborhood search to solve a two-tiered transportation problem, with buses transporting goods from a consolidation center to bus stops, followed by city freighters for last-mile delivery. \cite{Li2021} formulated a MILP model for an urban rail line to combine passenger and freight transportation, aiming to maximize the freight transportation profits using dedicated freight trains or existing passenger trains. \cite{Sahli2022} addressed the freight transportation scheduling problem using an urban passenger rail network. They proposed a genetic algorithm to minimize the total waiting time for parcel departures. 
\cite{Di2022} investigated the joint optimization of passengers and freight on metro-based systems and developed a Benders decomposition to find a near-optimal solution to address the train carriage arrangement and demand flow control. \cite{Sun2023} designed a three-tier metro-underground logistics service network, where they established a complex model that was solved by a particle swarm optimization algorithm for the hub location, customer allocation, and flow routing. \cite{schmidt2024using} studied the last-mile delivery problem with scheduled lines where public transport with a predefined timetable is applied to carry freight into the city center via dedicated bus stops. Through the exact and heuristic branch-price-and-cut algorithms, their results demonstrated that a dense network of lines and stops can significantly reduce routing costs and enable the completion of final deliveries with a reduced number of city freighters.

From a strategic planning perspective, \cite{Azcuy2021} demonstrated that utilizing public transport capacity for freight transfer can lead to significant savings, providing an efficient heuristic for determining the transfer station location where freight is later delivered by small vehicles to end-customers. 
In contrast, \cite{Kizil2023} focused on the expansion of service areas within public transport networks by exploring transfer opportunities to broaden service reach. They devised a branch-and-price method to address their specific challenges. \cite{Donne2023} introduced three mathematical formulations to support strategic decisions on which public transport lines and stop locations will be selected for service.

The prior review highlights the diversity of approaches and the growing interest in leveraging public transport for urban logistics. 


\subsection{Drone Delivery Problems}
Recently, drone delivery has become a revolutionary aspect of last-mile delivery, offering cost-effectiveness and environment-friendly solutions. 

\cite{Perera2020} investigated the application of drones for parcel delivery in scenarios where customers are uniformly distributed along a circular circumference. Their study indicated that as drone technology matures, associated costs are expected to decrease, leading to a more decentralized delivery network. While uncertain weather conditions can affect the flight time of drones to their destinations. To mitigate this, \cite{Cheng2024} proposed a robust model to optimize the drone schedule to effectively reduce service lateness at customers for drone delivery systems. 

However, drone delivery services are quite limited by the flight range and carrying capacity. Existing literature presents distinct approaches to address these challenges through vehicle coordination. \cite{Carlsson2018} proposed a novel configuration where a truck functions exclusively as a mobile depot, enabling a drone to perform sequential deliveries by periodically returning to the moving vehicle for package reloading. In contrast, ongoing research centers on coordinating drones with trucks for demand fulfillment. \cite{Murray2015} studied the scenario of a truck working collaboratively with drones for delivery, and proposed a heuristic to minimize the makespan. Building on this, \cite{Agatz2018} introduced an IP formulation and a two-stage heuristic to achieve better solutions. \cite{Wang2019} extended the model to multiple trucks, where each drone can travel with a truck, take off from its stop to deliver parcels, and land at a docking hub to switch to another truck while respecting the loading capacity and flight duration constraints.
\cite{Pugliese2021} addressed the delivery problem of trucks equipped with drones under time window, capacity, and flight duration constraints. They only considered time window constraints for the truck service because drones can deliver parcels without the customer being present. \cite{Madani2024} investigated the tandem operation of a truck and a drone, considering that the truck can launch and retrieve the drone from both customer and non-customer nodes, while the drone performs multiple visits per dispatch. They developed a variable neighborhood search-based algorithm to address this problem, and the effectiveness of the proposed model was demonstrated from the perspective of cost savings. 

Other than using trucks as mobile carriers of drones, \cite{Choudhury2021} considered the possibility of drones landing on public transit vehicles to travel on them, conserving energy and extending service coverage. Their model aims to minimize the makespan in delivering all packages from depots to customers. \cite{Moadab2022} deployed carbon-free technologies for drone charging on public transportation and minimized the total energy consumption in delivery operations. \cite{Cheng2023} developed an adaptive large neighborhood search algorithm for the simultaneous transportation of passengers and freight using demand-responsive buses and drones. The authors proposed that drones can perform parcel delivery by taking off from and landing on the buses’ rooftops automatically.  

Drone delivery can also cooperate with a fixed location called a parcel locker that provides space for loading and recharging. \cite{Zou2023} explored an integrated system of parcel lockers and drones where parcels are delivered from the lockers to customers by drone. They developed a genetic algorithm to minimize annual operating costs by determining the location of the lockers, the assignment of customer demands to each locker, and the number of drones at each locker. With a high degree of automation between lockers and drones, \cite{Zhu2024} developed a two-stage method for a hybrid problem that plans feasible routing and parking for drones and lockers such that the total travel distance is minimized.

This paper introduces a paradigm wherein buses first transport parcels to lockers at bus stops, and the drones positioned at bus stops deliver the parcels to customers. The integration of these elements into a cohesive system presents a difficult challenge due to the complex interaction among the included elements. To the best of our knowledge, this is the first paper to study this problem and its solution approaches (both exact and heuristic) addressing the freight multimodal transport system with buses and drones.

\subsection{Solution Approaches}
Solution approaches based on column generation (branch-and-price, branch-cut-and-price, etc.) are proven techniques for large and hard integer programming problems \citep{barnhart1998branch, Desaulniers2005}. This is especially the case for routing problems, where the current state-of-the-art exact methods are all found in this category \citep{Costa2019,florio2020new, pessoa2020generic, rostami2021branch}. However, our approach  integrates column generation and Benders decomposition. Even thought some studies, such as these by \cite{Restrepo2018}, \cite{Karsten2018}, and \cite{Daryalal2023}, consider this combination, the distinct column structures inherent in our problem necessitate the use of column generation in both master problem and sub-problems. To address the integer requirements in both stages, we employ branching strategies to ensure the derivation of an exact optimal solution.


\section{Problem Description and Formulation} \label{secProblem}
This section begins with an overview of the considered system in Section \ref{secSystem}, followed by a formal introduction of the notations and assumptions in the optimization problem in Section \ref{secAssumption}. We end this section with a compact MILP model in Section \ref{secFormulation}.

\subsection{Bus-Drone Freight Multimodal Transport System} \label{secSystem}
This paper introduces a bus-drone freight multimodal transport system aimed at enhancing logistics efficiency and sustainability. The system, as illustrated in Figure~\ref{figIntro}, is organized into three key stages: first-mile consolidation, bus transportation, and last-mile delivery.

{\bf Stage 1: First-Mile Consolidation.}
The initial stage encompasses the consolidation of parcels at the bus terminal. In urban logistics, parcels are often initially scattered and uncertain in location. E-commerce companies typically gather these parcels at a distribution center before proceeding with deliveries. For seamless integration with bus-based delivery services, it is advantageous for the distribution center to be close to the bus terminal, facilitating efficient loading and unloading processes. In this scenario, the bus terminal also acts as a temporary warehouse. Similar applications also exist in reality \citep{Masson2017, Kunming2022}. Generally, companies transport parcels from the distribution center to the terminal using full truckloads, following the shortest path. Therefore, this stage is considered direct transportation with a predetermined time and cost, allowing us to skip the freight consolidation process in the first-mile delivery and focus primarily on the subsequent stages.

{\bf Stage 2: Bus Transportation.}
During the bus transportation phase, parcels are conveyed on designated buses along a fixed bus line, stopping at designated bus stops where they are loaded into parcel lockers by the bus driver. We take the known schedules of the bus line as input. However, larger freight volumes could increase the unloading times for bus drivers, leading to an additional increase in the scheduled timetable's dwell time. Because this schedule is publicly known, any modification may deteriorate the service quality for passengers. Consequently, the operation time is the key to the system. Following the examples set by \cite{Masson2017}, we assume that all customer demands are packed into standard-sized parcel boxes, limiting the number of boxes that can be unloaded each time at a stop. This ensures the ease of carrying and stacking simultaneously without any operational difficulty to the driver. As such, the dwell time specified in the timetable is also assumed to be enough for the operation and is negligible when compared to the travel time between two bus stops.

{\bf Stage 3: Last-Mile Delivery.}
The final phase describes the delivery of parcels to end-customers, the ultimate destination in the field of e-commerce logistics. Parcels assigned for bus transport are stored in lockers at bus stops, where the advent of parcel lockers provides efficient storage for freight during distribution. Traditionally, the designated personnel will pick up the parcels and complete the delivery throughout the region. In contrast, we propose an environment-friendly mode where these lockers are equipped with drones. Each drone is capable of making multiple single-delivery round trips within its load capacity and flight duration, delivering parcels directly to customers.

\subsection{Notations and Assumptions} \label{secAssumption}
We consider a scenario where multiple buses on a single bus line deliver parcels to parcel lockers at bus stops, which are subsequently delivered to customers by drones. The terminal, denoted by $o$, consolidates all parcels from a set of customers $\mcC$. The demand of each customer $i \in \mcC$ is packed into a box, specifying a volume $q_i$ that is indivisible and must be delivered by a specified deadline $l_i$. For notational consistency, let $q_o = 0$ and $[0, l_o]$ represent the planning horizon. 

Operating on a predetermined timetable, a fleet of buses $\mcB$ depart from the terminal in sequence, traversing a set of bus stops $\mcS$. Travel times between consecutive stops are considered known and fixed, with road conditions during off-peak hours disregarded. The arrival time of bus $b \in \mcB$ at stop $s \in \mcS$ is denoted as $e_{bs}$. Note that $e_{b_2s} > e_{b_1s}$ if $b_2 > b_1$. Each bus $b$ possesses a capacity $Q^B$ for parcel transportation, ensuring that passenger satisfaction is not compromised. Parcel boxes are assigned to buses and transported to the stops for unloading. To avoid extended dwell times, assume that each bus unloads at most $Q^S$ boxes at a stop simultaneously and operates for a fixed duration of $\tau^B$, including the dwell time.   
 
At each stop $s$, a parcel locker is installed for storage purposes. The notation $\mcS$ is conveniently used to represent both the stops and their associated lockers. We assume that the lockers have sufficient capacity to accommodate the assigned parcels, based on the operator's policy of providing locker services only when there is available space to receive and store parcels \citep{dosSantos2022}. However, we consider a holding cost, $f^H$, at each locker, which is assessed per unit of time and per unit of freight. The parcels are loaded onto the designated drone according to the schedule in the locker. A set of drones $\mcD_s$ at locker $s$ collect parcels from potentially different buses and deliver them to customers. The time required for a drone to serve a customer is represented by $\tau^D$. As drones can execute deliveries by landing and releasing parcels on a customer's balcony, even in his/her absence, we focus on a deadline constraint instead of a time window \citep{Pugliese2021}. 

The drones are identical and denoted by a set $\mcD = \cup_{s \in \mcS} \mcD_s$. Each drone $d \in \mcD$ can perform multiple round trips, launching from and returning to the same locker. Generally, it is assumed that the drone can carry any customer demand in a single trip. However, drone deliveries are typically subject to constraints such as the flight duration and battery capacity \citep{Cheng2023}. Therefore, we assume that each drone can deliver a single demand within a specific radius from the locker on each trip. Let $\mcC_s$ be the subset of customers reachable by drones originating from locker $s$. Before embarking on a new trip, a drone requires a constant time $\tau^S$ to load boxes and swap a fully-charged battery. An ample number of batteries is assumed to be available, as in \cite{Poikonen2019}, and they can be recharged by the locker during periods of inactivity \citep{Zou2023}. To prevent prolonged operation, each drone has a maximum operation duration of $\Delta$, facilitating daily maintenance. We also assume that we have sufficient drones, but we want to make the best economical use of each drone.
	
The \prob\ is defined on a graph $G = (\mcV, \mcA)$, where $V = \{o\} \cup \mcS \cup \mcC$ represents the vertex set, and $\mcA$ denotes the arc set. The arc set $\mcA$ consists of both bidirectional arcs $(s,i)$ and $(i,s)$ that can be traveled by drones between each stop $s \in \mcS$ and the corresponding reachable customer $i \in \mcC_s$. Each arc $(i,j) \in \mcA$ is associated with a travel time $\tau_{ij}$. Consequently, the total operation duration for a drone's round trip from stop $s$ to customer $i$ is calculated as $\delta_{si} = \tau^S + \tau_{si} + \tau^D + \tau_{is}$. We use $c_{si}^2$ to denote the operation cost for this round trip, which includes the battery and parcel box operations. Note that $\mcC_s$ must also consider the customers' deadlines. Accordingly, we define $\mcB_s(i) = \{b \in \mcB| e_{bs} + \tau^B + \tau^S + \tau_{si} \leq l_i, i \in \mcC_s\}$ as the set of buses capable of fulfilling demand $i$ to stop $s$. 

Because customers are not pre-assigned to buses or stops, the problem involves determining the assignment schemes for buses and the delivery sequences for drones. Let $c_{bs}^1$ denote the operation cost per unit of freight on bus $b$ to stop $s$, including costs for loading, transportation, and unloading.  Let $f^F$ denote the fixed cost per drone, covering daily maintenance and purchase expenses. The objective is to minimize total costs, consisting of the operation cost of buses ($c_{bs}^1$), holding cost of lockers ($f^H$), and the fixed ($f^F$) and operation costs ($c_{si}^2$) of drones, while ensuring that each customer is served exactly once by a drone before his/her deadline. Note that we summarize the key parameters and decision variables in Table~\ref{tabNotations}. 



Before proceeding, we discuss the hardness of the \prob. When parcel assignments to buses and stops during the bus transportation stage are neglected, the \prob\ can be reduced to the multiple depot vehicle routing problem, which is known to be NP-hard \citep{Stodola2024}. As a result, the \prob\ is NP-hard. We formally state this result in the following proposition.
\begin{proposition}
	The \prob\ is NP-hard.
\end{proposition}


\subsection{Compact MILP Model} \label{secFormulation}

We present a compact MILP formulation. Define three sets of binary variables. The first one is $x_{sib}$, which is one if bus $b$ delivers demand $i$ to stop $s$, and zero otherwise. The second one is $y_{ijsd}$, which is one if drone $d$ associated with locker $s$ serves customer $i$ before customer $j$, and zero otherwise. Note that $i$ (resp. $j$) can be the station when $j$ (resp. $i$) is the first (resp. last) customer. The third one is $z_{sd}$, which is one, if drone $d$ is launched from locker $s$, and zero otherwise. Define two sets of continuous variables. We have $h_i$ for the holding time of the parcel for customer $i$ and $w_i$ for the service start time of the drone at customer $i$. The \prob\ is defined by the following compact MILP formulation, which is referred to as MILP1:
\begin{align}
\mbox{[MILP1]}~\min~
&\sum_{b \in \mcB}\sum_{s \in \mcS}\sum_{i \in \mcC_s} (c_{bs}^1 q_i + c_{si}^2) x_{sib} 
+ f^F \sum_{s \in \mcS}\sum_{d \in \mcD_s} z_{sd} 
+ f^H \sum_{i \in \mcC} q_i h_i \label{MF:obj}
\end{align}
Objective function (\ref{MF:obj}) minimizes the sum of the operation costs of buses and drones, the fixed cost of drones, and the holding cost of lockers. It is subject to the following constraints. 

The parcel assignments to buses and stops are presented as follows:
\begin{align}
&\sum_{b \in \mcB}\sum_{s \in \mcS} x_{sib} = 1, ~\forall~ i \in \mcC, \label{MF:visit}\\
&\sum_{s \in \mcS}\sum_{i \in \mcC_s} q_i x_{sib} \leq Q^B, ~\forall~ b \in \mcB, \label{MF:busCap}\\
&\sum_{i \in \mcC_s} q_i x_{sib} \leq Q^S, ~\forall~ b \in \mcB, s \in \mcS, \label{MF:stopCap}\\
&\sum_{d \in \mcD_s}\sum_{j \in \{s\} \cup \mcC_s} y_{jisd} = \sum_{d \in \mcD_s}\sum_{j \in \{s\} \cup \mcC_s} y_{ijsd} = \sum_{b \in \mcB} x_{sib}, ~\forall~ s \in \mcS, i \in \mcC_s, \label{MF:connect}
\end{align}
Constraints (\ref{MF:visit}) guarantee that each customer is served exactly once by a bus and a drone. Constraints (\ref{MF:busCap}) and (\ref{MF:stopCap}) restrict the number of boxes that can be loaded onto the bus and unloaded at the stop, respectively. Constraints (\ref{MF:connect}) link the decision variables $\boldsymbol{x}$ and $\boldsymbol{y}$ by assigning one predecessor and one successor for each demand when it is carried by bus $b$ to stop $s$. 

Drone flight constraints include the following:
\begin{align}
&\sum_{i \in \mcC_s} y_{sisd} = \sum_{i \in \mcC_s} y_{issd} \leq z_{sd}, ~\forall~ s \in \mcS, d \in \mcD_s, \label{MF:samelocker}\\
&\sum_{j \in \{s\} \cup \mcC_s} y_{jisd} - \sum_{j \in \{s\} \cup \mcC_s} y_{ijsd} = 0, ~\forall~ s \in \mcS, d \in \mcD_s, i \in \mcC_s, \label{MF:flow}\\
&\sum_{i \in \mcC_s}\sum_{j \in \{s\} \cup \mcC_s} \delta_{si} y_{ijsd} \leq \Delta z_{sd}, ~\forall~ s \in \mcS, d \in \mcD_s, \label{MF:duration}
\end{align}
Constraints (\ref{MF:samelocker}) ensure that the drone must launch from and return to the same locker. Constraints (\ref{MF:flow}) define the sequence of customers served by the same drone. Constraints (\ref{MF:duration}) impose the maximum operation duration on each drone. 

The subsequent expressions constitute the time-relationship constraints:
\begin{align}
&e_{bs} + \tau^B + \tau^S + \tau_{si} - w_i \leq M_{sib}(1 - x_{sib}), ~\forall~ b \in \mcB, s \in \mcS, i \in \mcC_s, \label{MF:startTime}\\
&w_i + \tau^D + \tau_{is} + \tau^S + \tau_{sj} - w_j \leq M_{ijs}(1 - \sum_{d \in \mcD_s} y_{ijsd}), ~\forall~ s \in \mcS, i \in \mcC_s, j \in \mcC_s, i\neq j, \label{MF:timeFlow}\\
&w_i - \tau_{si} - \tau^S - \tau^B - e_{bs} - h_i \leq M_{si}(1 - x_{sib}), ~\forall~ b \in \mcB, s \in \mcS, i \in \mcC_s, \label{MF:holding}\\
&w_i \leq l_i, ~\forall~ i \in \mcC, \label{MF:deadline}
\end{align}
Constraints (\ref{MF:startTime}) trace the time relationship between the buses and drones, specifically requiring that the service start time of the first customer is greater than the arrival time of the bus where the corresponding demand is unloaded. Constraints (\ref{MF:timeFlow}) express the service start time of customer $j$ when a drone visits it right after customer $i$. Constraints (\ref{MF:holding}) calculate the holding time of each demand at the locker. Note that three large constants are introduced for linearization, where $M_{sib} = e_{bs} + \tau^B + \tau^S + \tau_{si}$, $M_{ijs} =l_i + \tau^D + \tau_{is} + \tau^S + \tau_{sj}$, and $M_{si} = l_i - \tau_{si} - \tau^S - \tau^B - e_{bs}$. Constraints~(\ref{MF:deadline}) make sure the service start time is no later than the customer's deadline. 

Finally, constraints (\ref{MF:x}) -- (\ref{MF:w}) state the domain of all decision variables: 
\begin{align}
&x_{sib} \in \{0,1\}, ~\forall~ b \in \mcB, s \in \mcS, i \in \mcC_s, \label{MF:x}\\
&y_{ijsd} \in \{0,1\}, ~\forall~ s \in \mcS, d \in \mcD_s, i \in \{s\} \cup \mcC_s, j \in \{s\} \cup \mcC_s,\\
&z_{sd} \in \{0,1\}, ~\forall~ s \in \mcS, d \in \mcD_s,\\
&h_i \geq 0, ~\forall~ i \in \mcC,\\
&w_i \geq 0, ~\forall~ i \in \mcC. \label{MF:w}
\end{align} 

While MILP1 is valid for the \prob, it is not a computationally viable one (one reason is that the big-$M$ approach is used in constraints~\eqref{MF:startTime}--\eqref{MF:deadline}). It is well-known that the big-$M$ approach generally leads to weak lower bounds and does not provide satisfactory performance in practice. This motivates us to develop a more sophisticated and computationally viable approach for solving the \prob\ in the next section.


\section{Branch-Price-and-Benders-Cut} \label{secAlgorithm} 
This section first introduces an IP formulation with two sets of exponentially many variables (the two-column-based formulation) in Section \ref{secTCF} and then describes the Branch-Price-and-Benders-Cut solution method based on the IP formulation in the remaining part of this section.

\subsection{Two-Column-Based Formulation} \label{secTCF}
The two-column-based formulation allows for the implicit representation of constraints related to parcel assignments and drone flights through the definition of two sets of variables (i.e., columns), which can typically yield a better linear relaxation compared to a compact MILP \citep{Costa2019}. An assignment, denoted by $t$, is characterized by a set of customer parcels assigned to a bus and unloaded at a stop. Consequently, this set of parcels is served by the drones at this stop. An assignment is considered feasible if each customer is assigned at most once, and the number of boxes unloaded at the stop does not exceed $Q^S$. Let $\mcT$ be the set of all feasible assignments. We denote $\mcT_{bs}$ as the set of assignments associated with bus $b$ at stop $s$. Accordingly, $\mcT = \cup_{b \in \mcB}\cup_{s \in \mcS} \mcT_{bs}$. Let $\alpha_{it}$ be a binary coefficient that indicates whether customer $i$ is assigned to assignment $t$, and let $\beta_t$ be the total load in assignment $t$. The cost of each assignment $t$, including the operation costs of the bus and drones, is represented by $c_t^1$. 

Define a flight as an ordered set of round trips performed by a drone. A flight is deemed feasible if it respects all the assigned customers' deadlines and the maximum operation duration. Let $\mcU$, indexed by $u$, be the set of feasible flights. Denote by $\mcU_s$ the set of feasible flights associated with locker $s$. Accordingly, $\mcU = \cup_{s \in \mcS} \mcU_{s}$. We use a binary coefficient $\gamma_{ibu}$ to indicate whether flight $u$ serves customer $i$ unloaded by bus $b$. Each flight is associated with a cost $c_u^2$, which includes the fixed cost of the drone and the holding cost of the served freight.

Introducing two binary decision variables, $\theta_t$ and $\zeta_u$, which respectively represent whether assignment $t$ is selected and whether flight $u$ is selected, the two-column-based formulation, denoted as TCF, is presented as follows:
\begin{align}
	\mbox{[TCF]}~\min~
	&\sum_{t \in \mcT} c_t^1\theta_t + \sum_{u \in \mcU} c_u^2\zeta_u \label{TCF:Obj}\\
	\mbox{s.t.}~
	&\sum_{t \in \mcT} \alpha_{it}\theta_t = 1, ~\forall~ i \in \mcC, \label{TCF:visit}\\
	&\sum_{t \in T_{bs}} \theta_t \leq 1, ~\forall~ b \in \mcB, s \in \mcS, \label{TCF:BusStop}\\
	&\sum_{s \in \mcS}\sum_{t \in \mcT_{bs}} \beta_t \theta_t \leq Q^B, ~\forall~ b \in \mcB, \label{TCF:Capacity}\\
	&\sum_{u \in \mcU_s} \gamma_{ibu} \zeta_u = \sum_{t \in \mcT_{bs}} \alpha_{it}\theta_t, ~\forall~ s \in \mcS, i \in \mcC_s, b \in \mcB_s(i), \label{TCF:coupling}\\
	&\theta_t \in \{0,1\}, ~\forall~ t \in \mcT, \label{TCF:theta}\\
	&\zeta_u \in \{0,1\}, ~\forall~ u \in \mcU. \label{TCF:zeta}
\end{align}
Objective function (\ref{TCF:Obj}) aims to minimize the total costs of the integrated system. Constraints (\ref{TCF:visit}) ensure that each customer is assigned exactly once to a bus and then served by a drone. Constraints (\ref{TCF:BusStop}) enforce the selection of at most one assignment associated with bus $b$ at stop $s$. Constraints (\ref{TCF:Capacity}) respect the capacity of each bus. Constraints (\ref{TCF:coupling}) are coupling constraints that require drones to serve customers assigned to the corresponding bus. Finally, constraints (\ref{TCF:theta}) and (\ref{TCF:zeta}) define the decision variables as being binary. Given that we are minimizing the total cost, we can relax constraints (\ref{TCF:visit}) and (\ref{TCF:coupling}) to $\geq$ inequalities as they would be binding for an optimal solution.


Given the exponential number of assignments and flights, it is impractical to write out the whole formulation, except for very small instances. As a workaround, we apply column generation to TCF, avoiding the need for exhaustive enumeration \citep{Desaulniers2005}. This method initiates with a subset of assignments and flights, denoted as columns, to tackle the linear relaxation of TCF, referred to as the master problem. Thereafter, a pricing problem is solved for new variables with negative reduced costs, which, if found, are added to resolve the master problem (see Section \ref{secCG}). Moreover, we propose three branching strategies within a branch-and-bound framework (see Section~\ref{secBranch}). 

However, the coupling constraints (\ref{TCF:coupling}) lead to an astronomical number of potential combinations between assignments and flights. Every newly generated variable can interact with all existing ones, complicating the direct application of column generation. To overcome this difficulty, the key observation is that TCF can be decomposed into a two-stage decision problem. Specifically, in the first stage, parcel assignments to buses are determined based on the timetable, and the second stage focuses on scheduling drone visits for the parcels assigned to each locker, following this assignment scheme of the first stage. Moreover, each drone flight is exclusive to its associated locker, meaning that parcel deliveries among different lockers are independent. In other words, the second-stage problem can be decomposed into a set of subproblems by locker (bus stop). 
Guided by this insight, we apply Benders decomposition, an algorithm designed for partitioning large-scale two-stage models \citep{Naderi2021, Wu2023}. This method converts the relationship between $\boldsymbol{\theta}$ and $\boldsymbol{\zeta}$ into a set of Benders cuts, which allows the removal of constraints (\ref{TCF:coupling}) from TCF and uses Benders cuts for dynamic evaluation (see Section \ref{secBD}).

\begin{figure}[tb]
\vspace{-.4cm}
	\centering
\includegraphics[width = 0.95 \textwidth]{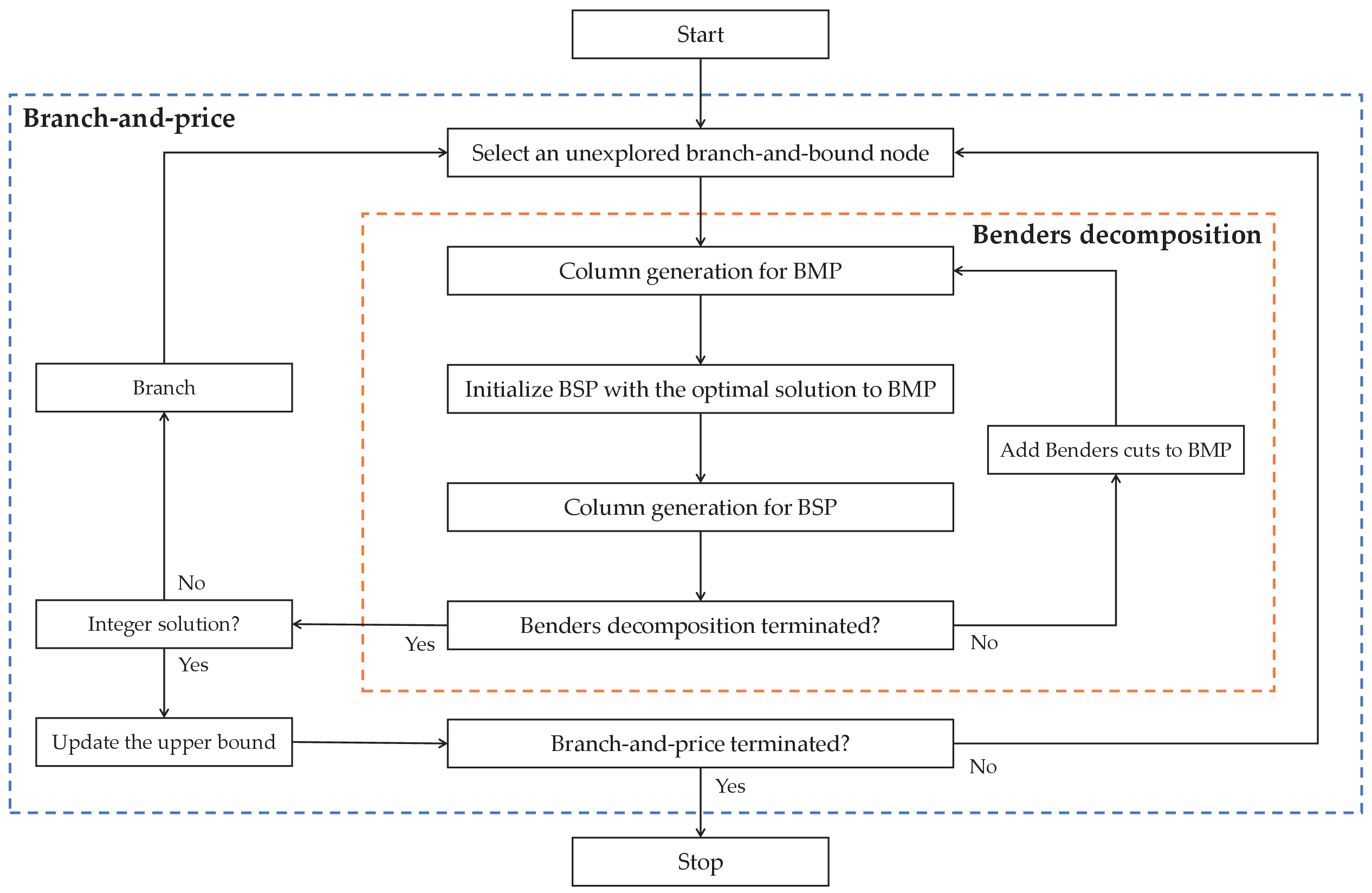}
	\caption{The Flowchart of the BPBC Algorithm.}	\label{figFlowchart}	
 \vspace{-.5cm}
\end{figure}

Following this logic, we propose a Branch-Price-and-Benders-Cut (BPBC) algorithm that integrates column generation and Benders decomposition within a branch-and-bound framework built on TCF, as depicted in Figure \ref{figFlowchart}.  In the subsequent sections, we delve into the specifics of the proposed algorithm.


\subsection{Benders Reformulation} \label{secBD}
This section elaborates on how to solve the linear relaxation of TCF at each branch-and-bound node. Given a feasible assignment scheme $\boldsymbol{\theta}$ respecting constraints (\ref{TCF:visit}) -- (\ref{TCF:Capacity}), the linear relaxation of TCF is projected onto the Benders subproblem (BSP), which can be further decomposed into $|S|$ subproblems. Each of these subproblems corresponds to a drone scheduling problem, denoted by BSP$_s$($\boldsymbol{\theta}$), and is formulated as follows:
\begin{align}
	\mbox{[BSP$_s$($\boldsymbol{\theta}$)]}~\min~&\sum_{u \in \mcU_s} c_u^2\zeta_u \label{BSP:obj}\\
	\mbox{s.t.}~&\sum_{u \in \mcU_s} \gamma_{ibu} \zeta_u \geq \sum_{t \in \mcT_{bs}} \alpha_{it}\theta_t, ~\forall~ i \in \mcC_s, b \in \mcB_s(i), \label{BSP:visit}\\
	&\zeta_u \geq 0, ~\forall~ u \in \mcU_s. \label{BSP:lb}
\end{align}
While BSP$_s$($\boldsymbol{\theta}$) still has exponentially many variables, we concentrate exclusively on non-zero parcel assignments (given parcel $i$, the corresponding $\alpha_{it}>0$ for some assignment $t$) that satisfy constraints (\ref{BSP:visit}), enabling the search for flights with negative reduced costs on a reduced network. Therefore, BSP$_s$($\boldsymbol{\theta}$) can be solved more efficiently.

Let $\omega_{ib}$ be the dual variables associated with constraints (\ref{BSP:visit}). The dual of the linear relaxation of BSP$_s$($\boldsymbol{\theta}$) is as follows:
\begin{align}
	\mbox{[DBSP$_s$($\boldsymbol{\theta}$)]}~\max~
	&\sum_{i \in \mcC_s}\sum_{b \in \mcB_s(i)}\sum_{t \in \mcT_{bs}} \alpha_{it}\theta_t \omega_{ib}\\
	\mbox{s.t.}~
	&\sum_{i \in \mcC_s}\sum_{b \in \mcB_s(i)} \gamma_{ibu} \omega_{ib} \leq c_u^2, ~\forall~ u \in \mcU_s, \label{DBSP:visit}\\
	&\omega_{ib} \geq 0, ~\forall~ i \in \mcC_s, b \in \mcB_s(i). \label{DBSP:omega}
\end{align}

\begin{theorem} \label{proDBSP}
	For any $s \in \mcS$, BSP$_s$($\boldsymbol{\theta}$) and DBSP$_s$($\boldsymbol{\theta}$) are feasible and their polyhedrons are bounded.
\end{theorem}

According to Theorem \ref{proDBSP}, let $\Omega_s$ be the set of extreme points in the polyhedron defined by constraints (\ref{DBSP:visit}) -- (\ref{DBSP:omega}); both BSP$_s$($\boldsymbol{\theta}$) and DBSP$_s$($\boldsymbol{\theta}$) have the same finite optimal value
\begin{equation}
	\max\limits_{\boldsymbol{\omega} \in \Omega_s} \bigg(\sum_{i \in \mcC_s}\sum_{b \in \mcB_s(i)}\sum_{t \in \mcT_{bs}} \alpha_{it}\theta_t \omega_{ib}\bigg).
\end{equation}

With a set of nonnegative continuous variables $\varphi_s$, we reformulate the linear relaxation of TCF, called the Benders master problem (BMP), as follows:
\begin{align}
	\mbox{[BMP]}~\min~
	&\sum_{t \in \mcT} c_t^1\theta_t + \sum_{s \in \mcS} \varphi_s \label{BMP:Obj}\\
	\mbox{s.t.}~&\mbox{(\ref{TCF:visit}) -- (\ref{TCF:Capacity}), } \nonumber\\
	&\varphi_s \geq \sum_{i \in \mcC_s}\sum_{b \in \mcB_s(i)}\sum_{t \in \mcT_{bs}} \alpha_{it}\omega_{ib}\theta_t, ~\forall~ s \in \mcS, \boldsymbol{\omega} \in \Omega_s, \label{BMP:ClassicCut}\\
	&\theta_t \geq 0, ~\forall~ t \in \mcT.
\end{align}
Enumerating the complete set of Benders cuts (\ref{BMP:ClassicCut}) is impractical due to their exponential size. They are added to BMP dynamically when the optimal value of BSP$_s$($\boldsymbol{\theta}$) exceeds $\varphi_s$ for some $s \in \mcS$. 

\subsection{Column Generation} \label{secCG}

The pricing subproblems of both BMP and BSP can be modeled as a variant of the elementary shortest path problem with resource constraints (ESPPRC), respectively. The ESPPRC is NP-hard in the strong sense \citep{dror1994note}. To identify new columns with negative reduced costs, we propose a labeling algorithm. This algorithm is a widely used dynamic programming approach to generate feasible complete paths from the source to the sink on the ESPPRC graph. It has shown remarkable efficacy in solving various pricing subproblems within the column generation method \citep{Wei2020, Yu2022}. We refer the details to \ref{detail-secCG}.

\subsection{Branching Strategies} \label{secBranch}
The column generation terminates when the optimal value of BSP equals $\sum_{s \in \mcS} \varphi_s$. At this point, we obtain an optimal solution $(\boldsymbol{\theta}, \boldsymbol{\zeta})$ to BMP and the associated BSP at the branch-and-bound node. If the solution is fractional for some assignments or flights, we branch the current node into two child nodes, which must remove the current solution and retain an optimal integer solution for future exploration. Three branching strategies are employed in their hierarchy, which are detailed below.  

First, let $\Pi_{si}^1 = \sum_{b \in \mcB_s(i)}\sum_{t \in \mcT_{bs}} \alpha_{it}\theta_t$ indicate whether parcel $i$ is assigned to stop $s$. If there exists a fractional $\Pi_{si}^1$, we force $\Pi_{si}^1 = 0$ by removing parcel $i$ from set $C_s$. On the second branch, we force $\Pi_{si}^1 = 1$ by ensuring that parcel $i$ is assigned to stop $s$. Secondly, let $\Psi_{ib}^1 = \sum_{s \in \mcS}\sum_{t \in \mcT_{bs}} \alpha_{it}\theta_t$ indicate whether parcel $i$ is assigned to bus $b$. If there exists a fractional $\Psi_{ib}^1$, we force $\Psi_{ib}^1 = 0$ by ensuring that bus $b$ cannot serve customer $i$ on the first branch. Conversely, we force $\Psi_{ib}^1 = 1$ by directly assigning parcel $i$ to bus $b$ on the other branch. Finally, let $\sigma_{iju}$ be a binary variable equal to one if flight $u$ has a consecutive visit from customer $i$ to customer  $j$ (note that the visit might include the locker). If $\Lambda_{ijs}^2 = \sum_{u \in \mcU_s} \sigma_{iju}\zeta_u$ for a locker $s$ is fractional, we force $\Lambda_{ijs}^2 = 0$ by forbidding $i$ to be served consecutively before $j$ in one child node. In the other, $\Lambda_{ijs}^2 = 1$ requires a drone to serve $i$ and $j$ consecutively.  

We impose these branching constraints only on the associated subproblems to generate new variables, preserving the structures of both BMP and BSP. When multiple values of any strategy are fractional, we select the one closest to 0.5. Furthermore, a best-first strategy is employed to select an unexplored branch-and-bound node for the next iteration. Specifically, the node with the minimum objective function value of the associated BMP is selected.

\section{Algorithmic Enhancements} \label{secImprovement}
This section presents several algorithmic enhancements to accelerate the convergence of the lower and upper bounds in the solution process. As will become evident later, these enhancements are crucial for the BPBC algorithm.

\subsection{Warm-Start Strategies} \label{secWS}
Warm start is an acceleration technique that leverages information from previous calculations to potentially reduce the iteration count and improve algorithm efficiency. At the root node, we warm start the BMP of the root node using a simple greedy heuristic to find an initial solution.

For any other branch-and-bound node, we propose two warm-start strategies for its BMP: one using variables and the other using Benders cuts. Firstly, the initial variables of each node are inherited from the set of variables with non-zero values from its parent node. Secondly, we utilize Benders cuts~(\ref{BMP:ClassicCut}) detected thus far for a warm start. It should be noted that cuts detected at a node with a set of branching constraints defined by variables $\Lambda_{ijs}^2$ are only valid for initializing nodes with a superset of these branching constraints. In this case, the BSP of a parent node is a relaxation for those of its child nodes. Further, Benders cuts~(\ref{BMP:ClassicCut}) are defined by the dual points of DBSP. Therefore, all such cuts from a parent node remain valid for its child nodes, as the DBSP polyhedrons of the child nodes can be projected onto that of the parent node. It follows that Benders cuts detected by the root node can serve as initial cuts for the BMP of any other node.


\subsection{Bounding-and-Fixing Strategies} \label{secBF}
Let $UB$ denote the current global upper bound for TCF. If a lower bound for BMP exceeds $UB$, we can prune the current node to avoid further iterations. The following proposition defines a valid lower bound when solving BMP to optimality at each column generation iteration.

\begin{proposition} \label{proBound}
The lower bound for BMP is computed as $LB = \Gamma + \sum_{s \in \mcS}\sum_{b \in \mcB: v_{bs} < 0} v_{bs}$, where $\Gamma$ is the optimal value of the linear relaxation of the current BMP with the $\theta$ variables and constraints~(\ref{BMP:ClassicCut}) generated so far, and $v_{bs}$ is the reduced cost of an optimal column found in MPS$_{bs}$.
\end{proposition}

In addition, we propose another strategy for further acceleration by reducing the times of processing MPS$_{bs}$. Assuming the existence of specific $s^* \in \mcS$ and $b^* \in \mcB$ such that $v_{b^*s^*}+LB \geq UB$, it means that no parcel assignment can improve $UB$. Thus, we fix $\sum_{t \in \mcT_{b^*s^*}} \theta_t = 0$. Consequently, we can skip all MPS$_{b^*s^*}$ in the subsequent iterations for the current node or any derived child nodes. 



\subsection{Valid Inequalities} \label{secLB}
Before the incorporation of Benders cuts into BMP, the model may suffer from poor relaxation since a specific component of objective function (\ref{TCF:Obj}) has been projected out. To overcome this challenge, we introduce valid inequalities to reflect the limited information on the values of $\mathbold{\varphi}$. Given that the fixed costs of drones can be estimated to approximate the cost of BSP, we define the following valid inequalities:
\begin{equation} \label{lowerbound}
	\varphi_s \geq f^F \frac{\sum_{i \in \mcC_s}\sum_{b \in \mcB_s(i)}\sum_{t \in \mcT_{bs}}\alpha_{it} \delta_{si} \theta_t}{\Delta}, ~\forall~ s \in \mcS.
\end{equation}

These inequalities provide a lower bound for each corresponding $\varphi_s$ by determining the minimum number of drones associated with locker $s$. They are used to initialize BMP and are updated as more new columns are identified. We show that initial cuts (\ref{lowerbound}) are valid in the following proposition.

\begin{proposition} \label{proLowerbound}
Inequalities (\ref{lowerbound}) are valid for BMP.
\end{proposition}

\subsection{Primal Heuristics} \label{secUB}
In the early stages of the branch-and-bound exploration, obtaining a high-quality upper bound $UB$ for TCF can significantly prune more unpromising nodes and enhance the methods described in Section \ref{secBF}. To this end, we employ two approaches to refine $UB$.

The first approach lies in the fact that if $(\boldsymbol{\theta}, \boldsymbol{\zeta})$ are in integral values during the column generation, they constitute a feasible solution to TCF with an upper bound $UB = Z_{BMP} - \sum_{s \in \mcS} \varphi_s + Z_{BSP}$. Here, $Z_{BMP}$ and $(\mathbold{\theta}, \mathbold{\varphi})$ represent the optimal value and the optimal solutions of the current BMP, respectively. Similarly, $Z_{BSP}$ and $\mathbold{\zeta}$ represent the optimal value and optimal solutions for the associated BSP, respectively.

Another approach involves applying an IP problem to find a feasible upper bound. Before making the branching decision, we include all variables $(\mathbold{\theta}, \mathbold{\zeta})$ identified at the current node into TCF and solve the resulting TCF as an IP. Typically, a commercial solver can swiftly solve the resulting TCF or at least quickly obtain a feasible solution. We limit the execution time of this heuristic to 30 seconds per run.

\section{Computational Studies} \label{secExperiments}
This section first describes the test instances and experimental setup in Section \ref{secInstance}. Subsequently, we present the overall performance of the BPBC algorithm and discuss the impacts of the algorithmic enhancements on the BPBC algorithm in Section \ref{secPerformance}. Then, Section~\ref{secHeuristic} discusses a heuristic version of the BPBC algorithm and applies it to large instances. Finally, Section \ref{secSensitivity} provides sensitivity analyses and managerial insights based on the computational experiments.

\subsection{Test Instances and Experimental Setup} \label{secInstance}
\begin{figure}[tb]
\vspace{-.4cm}
	\centering
	\includegraphics[width = 0.55 \textwidth]{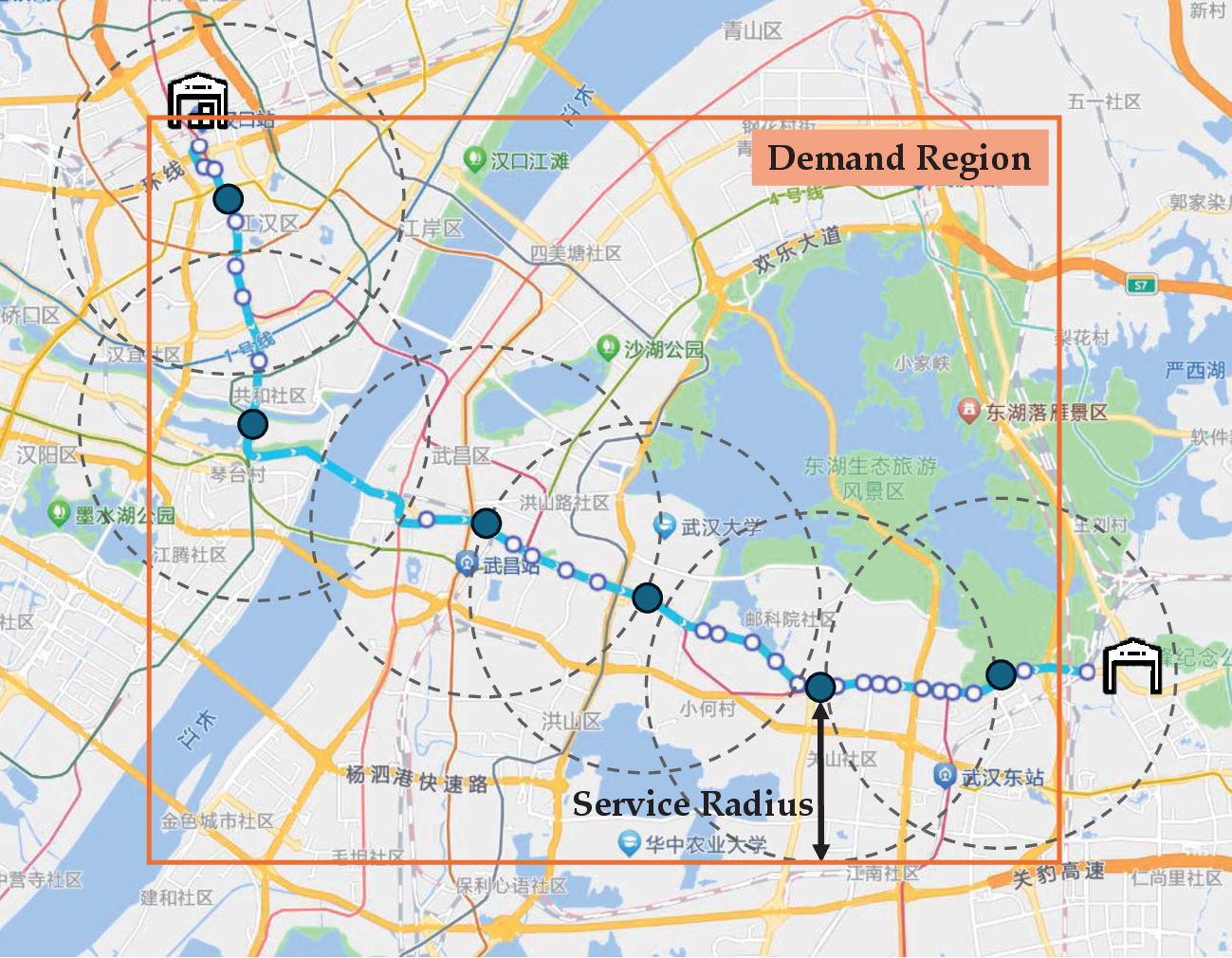}
	\caption{Illustration of the Test Instances (Generated by Baidu Maps).}	\label{figIns}	
 \vspace{-.5cm}

\end{figure}

Given the nascent stage of freight multimodal transport systems with buses and drones, we have generated synthetic instances based on real-world information and referenced parameters from the relevant literature.


For the creation of authentic spatial simulations, we examine the bus transportation network in Wuhan, China (the biggest city in the middle of China with a population of around 14 million inhabitants). The bus enterprise's timetable indicates that off-peak hours are from 10:00 to 16:00. We focus on a specific bus line that passes through 32 bus stops, taking approximately 93 minutes per trip. Within this 6-hour interval, 31 bus trips are scheduled for this line. We select six or eight buses from these trips for freight transportation, ensuring a satisfactory level of passenger service.

In constructing the delivery network, we embrace a logical principle: parcel lockers should be strategically dispersed to guarantee coverage of different areas and facilitate the monitoring of the drone delivery process. Consequently, we select six or eight bus stops along the line to establish lockers and deploy drones. Drones are assumed capable of performing a round trip within an 8 km radius, as suggested by Amazon \citep{Gross2013}. The scheduled arrival times of buses from the terminal to respective stops are extracted directly from the bus timetable. 

The logistics company consolidates parcels at the terminal for delivery to the end-customers within a distribution region. To ensure suitable distances among the terminal, stops, and end-customers during the network generation phase, we focus on a rectangular region of 20 by 16 square kilometers, centered on the bus line, as shown in Figure \ref{figIns}. The geographical locations of customers are uniformly distributed over this region and do not overlap with any selected bus stop. Furthermore, each customer is ensured to be within at least one locker's reachable radius. 

During off-peak hours, there is assumed to be sufficient space for the bus to carry parcels. We set the bus capacity for parcels at 50 kg, comparable to the weight of an average adult passenger. The maximum number of boxes that can be unloaded at each stop is set to 10 kg. Amazon's drones are capable of carrying up to 2.27 kg, while Workhorse claims that their drone can carry up to 4.54 kg \citep{Sacramento2019}. Based on these limits, we generate customer demands ranging from 0.5 kg to 4.5 kg in 0.5 kg intervals using a uniform distribution function. Additionally, we set the maximum operation duration of each drone to 120 minutes. The planning horizon extends from 10:00 to 18:00, with customer deadlines randomly distributed between these times, ensuring that at least one drone from a locker can meet each deadline. The parameters $\tau^B$, $\tau^S$, and $\tau^D$ are set to 1 minute, following the values used in \cite{Cheng2023}.

The travel distances between two points for drones are calculated using the Euclidean norm, with each drone maintaining an average speed of 40 km/h \citep{Wang2019}. The operation cost of drones is priced at \$2.0/km based on the electricity consumed per kilometer. We consider a fixed daily maintenance and purchase expenses of \$40 for drone usage. The holding cost for freight in lockers is assumed to be \$1.0 per kilogram per minute. The operation cost of delivering freight from the bus to the stop is \$1.0/km.

We generate two sets of instances. For the first set, each instance has six buses and six bus stops. We vary the number of customers with $|C| \in \{18, 27, 36, 45\}$.  Each setting consists of 10 instances with varied customer profiles to avoid biased data. There are 40 instances in total. For the second set, each instance has eight buses and eight bus stops. The number of customers is either 60 or 80. Five instances are generated for each setting. There are 10 instances in the second set. Experiments are conducted on an Intel Core i7-4720 with a 2.60 GHz CPU and 8 GB RAM, running Windows 11. The algorithms are coded in Java and utilize CPLEX 12.6. We set a maximum time limit of 3,600 seconds for any approach for each instance. For brevity, we report aggregated results for the first set of instances and provide detailed instance-level results in \ref{appDetailed}.
 
\subsection{Overall Performance of the BPBC Algorithm} \label{secPerformance}

This section presents the overall performance of the BPBC algorithm and analyzes the impacts of the algorithmic enhancements introduced in Section \ref{secImprovement}. While the compact formulation MILP1 can be directly solved by CPLEX, as expected, it does not perform well and reaches the time limit for most instances. Thus, we choose to omit the results of MILP1 and focus on the BPBC algorithm.

We consider five configurations that use the following abbreviations for ease of presentation:
\begin{itemize}
	\item Basic: The BPBC algorithm described in Section \ref{secAlgorithm} without any enhancement;
	\item Basic-W: Basic enhanced with the warm-start strategies described in Section \ref{secWS};
	\item Basic-WB: Basic-W enhanced with the bounding-and-fixing strategy described in Section \ref{secBF};
	\item Basic-WBI: Basic-WB enhanced with valid inequalities (\ref{lowerbound}) described in Section \ref{secLB};
	\item BPBC (Basic-WBIH): BPBC has all enhancements described in Section \ref{secImprovement}, enhancing BPBC-WBI with the primal heuristics in Section~\ref{secUB}.
\end{itemize}

\begin{table}[tb]
\TableSpaced
	\centering
	\caption{Impacts of the Algorithmic Enhancements.}
	\resizebox{\textwidth}{!}{
    \begin{tabular}{llllllllllllllll}
	\toprule
	&       & \multicolumn{2}{l}{Basic} &       & \multicolumn{2}{l}{Basic-W} &       & \multicolumn{2}{l}{Basic-WB} &       & \multicolumn{2}{l}{Basic-WBI} &       & \multicolumn{2}{l}{BPBC (Basic-WBIH)} \\
	\cmidrule{3-4}\cmidrule{6-7}\cmidrule{9-10}\cmidrule{12-13}\cmidrule{15-16}    Problem &       & Opt.  & CPU   &       & Opt.  & CPU   &       & Opt.  & CPU   &       & Opt.  & CPU   &       & Opt.  & CPU \\
	\midrule
Customer18 &    & 10    & 1.8   &       & 10    & 1.0   &       & 10    & 1.4   &       & 10    & 1.8   &       & 10    & 1.8  \\
Customer27 &    & 10    & 92.0  &       & 10    & 70.3  &       & 10    & 59.6  &       & 10    & 66.1  &       & 10    & 55.2  \\
Customer36 &    & 10    & 630.6  &       & 10    & 406.6  &       & 10    & 374.2  &       & 10    & 451.7  &       & 10    & 267.7  \\
Customer45 &    & 6     & 2598.2  &       & 7     & 2147.0  &       & 8     & 2021.5  &       & 10    & 1619.5  &       & 10    & 1353.1  \\
Average &       & 9.0   & 830.7  &       & 9.3   & 656.2  &       & 9.5   & 614.2  &       & 10.0  & 534.8  &       & 10.0  & 419.4  \\
	\bottomrule
\end{tabular}
	}
	\label{tabImprovements}
\end{table}

All five configurations are applied to the 40 instances in the first set. Table \ref{tabImprovements} reports the results with the following information: the number of instances solved to optimality (Opt.) and the average computational time in seconds (CPU), respectively.

An observation from Table \ref{tabImprovements} indicates that in Customer18 and Customer27, all configurations solve these 20 instances optimally within the time limit, although there are minor variations in computational times. This demonstrates the effectiveness of the proposed algorithm in solving small-scale experiments. However, the impacts of these enhancements become more pronounced as the problem size increases. Without any enhancement, the computational time for Basic increases substantially, for example, from 92.0 to 630.6 seconds for Customer36. Moreover, Basic cannot find four optimal solutions within the running time limit for Customer45. While Basic-W finds one more optimal solution and consumes less computational times for the instances. This emphasizes the importance of initializing columns and Benders cuts in BMP and shows the critical role of the warm-start strategies in managing large-scale experiments. The bound-and-fixing strategy can reduce the number of iterations and accelerate column generation, providing a better number of optimal solutions by Basic-WB than Basic-W, and shortening the computational time. Applying the initial cuts \eqref{lowerbound} improves the performance of Basic-WB by tightening the relaxation of BMP and solves all instances optimally. Finally, with the application of primal heuristics described in Section \ref{secUB}, high-quality upper and lower bounds are obtained for further improving the computational time.

In summary, the BPBP algorithm is able to provide satisfactory performance. Moreover, these enhancements can improve the solution process of the BPBC algorithm, offering advantages in solution speed and the number of optimal solutions obtained.





Next, we demonstrate the benefits of jointly optimizing the parcel assignments and drone flights by comparing the results of BPBC algorithm with those obtained using the sequential optimization (SO) algorithm. In the latter approach, the bus operation problem is first solved to determine the optimal parcel assignment, which is then used to sequentially optimize the drone operation and derive the corresponding optimal scheduling. The analysis encompasses all 40 instances in the first set, and the algorithmic enhancements have been adjusted to align with the SO algorithm.  

\begin{table}[tb]
\TableSpaced
	\centering
	\caption{Analysis of the BPBC and SO Algorithms.}
	\begin{tabular}{lllllllll}
		\toprule
		&       & \multicolumn{3}{l}{Cost} &       & \multicolumn{3}{l}{CPU} \\
		\cmidrule{3-5}\cmidrule{7-9}    Problem &       & BPBC  & SO & Saving &       & BPBC  & SO & Ratio \\
		\midrule
		Customer18 &       & 2302.05  & 2368.56  & 2.81\% &       & 1.8   & 0.18  & 10.03  \\
		Customer27 &       & 3396.57  & 3504.50  & 3.08\% &       & 55.2  & 0.18  & 313.80  \\
		Customer36 &       & 4186.59  & 4279.07  & 2.16\% &       & 267.7  & 0.18  & 1067.24  \\
		Customer45 &       & 5572.07  & 5639.22  & 1.19\% &       & 1353.1  & 0.20  & 6565.71  \\
		Average &       & 3864.32  & 3947.84  & 2.31\% &       & 419.5  & 0.19  & 2138.06  \\
		\bottomrule
	\end{tabular}
	\label{tabSequential}
\end{table}

Table \ref{tabSequential} gives the results in terms of the cost and computational time, where the columns ``Saving'' and ``Ratio'' denote the percentage of cost savings achieved by joint optimization and the ratio of the computational time, respectively. The SO algorithm efficiently solves all test instances with an average computational time of 0.19 seconds. In this approach, the optimization of drone scheduling is based solely on the parcel assignments made in the first stage, without fully considering the drone costs and holding costs that could affect the overall objective. Conversely, BPBC algorithm spends a longer time in solving the \prob\ because it seeks to balance the parcel assignments for bus stops and the routing of drone deliveries. 


On the other hand, given that SO overlooks the drone scheduling in the first stage, it generates a less favorable parcel assignment for drones, leading to increased costs. Compared to the SO, BPBC achieves an average reduction of 2.31\% in total costs, indicating its potential to yield significant benefits for logistics companies in making optimal decisions. Finally, the cost saving is not significantly correlated with the problem size, suggesting that SO can serve as a heuristic for larger-scale decision-making.

\subsection{Heuristic Version of the BPBC Algorithm on Large Instances}\label{secHeuristic}

This section provides further analysis of a heuristic version of the BPBC algorithm (HBPBC), designed to obtain high-quality solutions for instances with larger problem sizes. Specifically, HBPBC only executes the BPBC algorithm at the root node. Once the BMP at the root node is solved, we feed all generated columns to TCF and solve the resulting TCF as an IP problem.

\begin{table}[tb]
\TableSpaced
  \centering
  \caption{Comparative Results of BPBC and HBPBC on Customer45.}
    \begin{tabular}{llllllll}
    \toprule
&       &   \multicolumn{2}{l}{BPBC} &       &  \multicolumn{3}{l}{HBPBC} \\
            \cmidrule{3-4}                      \cmidrule{6-8}    
Instance &       & Cost  & CPU   &       & Cost  & CPU   & Gap \\
    \midrule
    Customer45-0 &       & 5226.25  & 1056.8  &       & 5315.95  & 15.7  & 1.69\% \\
    Customer45-1 &       & 5471.45  & 538.9  &       & 5516.85  & 361.5  & 0.82\% \\
    Customer45-2 &       & 6205.10  & 783.5  &       & 6241.50  & 58.6  & 0.58\% \\
    Customer45-3 &       & 5663.10  & 539.5  &       & 5706.15  & 9.8   & 0.75\% \\
    Customer45-4 &       & 5068.45  & 1354.4  &       & 5121.10  & 198.1  & 1.03\% \\
    Customer45-5 &       & 4935.55  & 1352.9  &       & 4940.45  & 46.9  & 0.10\% \\
    Customer45-6 &       & 5088.65  & 1238.5  &       & 5135.85  & 599.3  & 0.92\% \\
    Customer45-7 &       & 5863.45  & 998.4  &       & 5869.00  & 13.5  & 0.09\% \\
    Customer45-8 &       & 5598.85  & 2655.7  &       & 5603.10  & 28.1  & 0.08\% \\
    Customer45-9 &       & 6599.85  & 3012.8  &       & 6611.70  & 35.3  & 0.18\% \\
    \bottomrule
    \end{tabular}
  \label{tabCustomer45}
\end{table}
We first evaluate the performance of HBPBC on the 10 instances with 45 customers in the first set of instances. To evaluate the solution quality, we compare the solutions obtained by HBCBP to the optimal ones found by BCBP. Table~\ref{tabCustomer45} has the results. Among the 10 instances, the average runtime of HBPBC is 136.7 seconds. More importantly, the average, minimum, and maximum gaps of HBPBC are 0.62\%, 0.08\%, and 1.69\%, respectively. This provides us with some assurance regarding the quality of the solutions obtained by HBCBP.


Next, we conduct experiments on the second data set, which has 10 larger instances with 60 or 80 customers. Two benchmark solution approaches are employed to assess performance: an exhaustive attempt by CPLEX to solve MILP1 instances exactly, despite the impracticality of achieving optimality within a reasonable time, and the SO algorithm introduced in the previous section. We compare the solution quality by using the best solutions found within the time limit. 

\begin{table}[tb]
\TableSpaced
\centering
\caption{Analysis of the Heuristic Version of the BPBC Algorithm.}
\begin{tabular}{llllllllllll}
    \toprule
    Instance &       & HBPBC  & CPU   &       & MILP1 & CPU   & Gap$_{M}$ &       & SO    & CPU   & Gap$_{S}$\\
    \midrule
    Customer60-0 &       & 7073.55  & 323.0  &       & 7411.85  & 3600.0  & 4.56\% &       & 7251.85  & 1.9   & 2.46\% \\
    Customer60-1 &       & 6871.25  & 436.7  &       & 7287.15  & 3600.0  & 5.71\% &       & 7083.95  & 3.2   & 3.00\% \\
    Customer60-2 &       & 5870.80  & 442.5  &       & 6290.45  & 3600.0  & 6.67\% &       & 6259.90  & 5.0   & 6.22\% \\
    Customer60-3 &       & 7324.20  & 1152.2  &       & 7529.30  & 3600.0  & 2.72\% &       & 7611.90  & 1.4   & 3.78\% \\
    Customer60-4 &       & 7074.95  & 506.3  &       & 7432.75  & 3600.0  & 4.81\% &       & 7146.45  & 2.0   & 1.00\% \\
    Customer80-0 &       & 8654.90  & 681.6  &       & 10842.65  & 3600.0  & 20.18\% &       & 8849.15  & 5.5   & 2.20\% \\
    Customer80-1 &       & 7186.80  & 659.9  &       & 8425.70  & 3600.0  & 14.70\% &       & 7440.50  & 15.1  & 3.41\% \\
    Customer80-2 &       & 9117.85  & 1798.3  &       & 10813.80  & 3600.0  & 15.68\% &       & 9439.85  & 0.9   & 3.41\% \\
    Customer80-3 &       & 8310.80  & 882.2  &       & 10036.60  & 3600.0  & 17.20\% &       & 8662.15  & 3.9   & 4.06\% \\
    Customer80-4 &       & 8315.55  & 943.1  &       & 9944.10  & 3600.0  & 16.38\% &       & 8484.60  & 13.6  & 1.99\% \\
    Average &       &       & 782.0  &       &       & 3600.0  & 10.86\% &       &       & 5.2   & 3.16\% \\
    \bottomrule
	\end{tabular}
	\label{tabHeuristic}
\end{table}

Table \ref{tabHeuristic} reports the objective values of the solutions obtained by the three approaches and the gaps between HBPBC and the other two solution approaches, Gap$_{M}$ for MILP1 and Gap$_{S}$ for the SO algorithm. First of all, MILP1 not only fails to provide optimal solutions within the time limit, but also ends with bad quality solutions. However, HBPBC is still able to output decent solutions, which could be over 20\% better than the ones found by MILP1. On average, the gap between HBPBC and MILP1 (Gap$_{M}$) is 10.86\%. Compared to SO, HBPBC consistently obtains better solutions. The gap could be as large as 6.22\%, and the average gap is 3.16\%. Thus, HBPBC performs better in terms of the solution gap, albeit at the cost of increased computational time. 

In conclusion, the evidence from these comparisons strongly supports the effectiveness of HBPBC, positioning it as a viable solution strategy for real-world applications. Building on HBPBC, we might execute BPBC for better results if extra computational resources are available.



\subsection{Sensitivity Analyses} \label{secSensitivity}
To obtain deeper managerial insights on the integrated system, this section performs sensitivity analyses on various scenarios. First, we evaluate the environmental advantages of integrating buses and drones by comparing it to traditional logistics. Next, we discuss the impact of cost parameters on the \prob. Finally, we investigate the impact of the parcel locker configuration on the performance of the bus-drone system.

\subsubsection{Analysis of the Environmental Advantages:}

This section derives insights on the environmental advantages of integrating buses and drones by comparing it to traditional logistics. In the case where all customers must be served, we compare the carbon emissions for (i) buses and drones, and (ii) a single truck that starts and ends at the terminal. For the latter case, we roughly disregard the customer deadlines and use the Euclidean distance between two nodes, leading to an underestimation of its carbon emissions.

To ascertain the carbon emissions for these means of transport, we refer to the parameters provided by \cite{Eggleston2006} and \cite{Goodchild2018}. The average emission factor for diesel is 2.5479 kgCO$_2$/L, with buses consuming approximately 0.3L/km. Given that the average off-peak speed of buses is 15 km/h, the carbon emissions for buses are calculated by 2.5479 kgCO$_2$/L $\times$ the total diesel consumption while carrying parcels. For drones, carbon emissions are determined by the formula 0.3773 kgCO$_2$/kWh $\times$ 0.1 kWh/mile $\times$ total distance in miles, where 0.3773 kgCO$_2$/kWh represents the average carbon emission factor at power generation facilities consumed by a drone, and 0.1 kWh/mile represents the average energy requirement for delivery. The truck's speed is standardized at 40 km/h, which is about 25 mph when converted to miles per hour. The corresponding average emission factor is 1.2603 kgCO$_2$ per mile. Consequently, the carbon emissions for trucks are calculated by 1.2603 kgCO$_2$/mile $\times$ total distance in miles.

\begin{table}[htbp]
\TableSpaced
\centering
	\caption{Analysis of the Emissions (in kg of CO$_2$) for Different Means}
\begin{tabular}{lllllllllll}
	\toprule
	Problem &       & Bus-drone &       & Drone &       & Truck &       & Dev$_{BT}$ &       & Dev$_{DT}$ \\
	\midrule
	Customer18 &       & 64.5  &       & 0.5   &       & 51.6  &       & 12.9  &       & -51.1  \\
	Customer27 &       & 69.3  &       & 0.7   &       & 55.6  &       & 13.7  &       & -55.0 \\
	Customer36 &       & 72.4  &       & 0.8   &       & 60.8  &       & 11.5  &       & -60.0 \\
	Customer45 &       & 73.7  &       & 0.9   &       & 65.1  &       & 8.6  &       &  -64.2 \\
	Average &       & 70.0  &       & 0.7   &       & 58.3  &       & 11.7  &       & -57.6  \\
	\bottomrule
\end{tabular}
	\label{tabCarbon}
\end{table}

Table \ref{tabCarbon} presents a comparison of carbon emissions for different means. The ``Bus-drone'' and ``Drone'' columns respectively report the total emissions for the \prob\ and the emissions attributable solely to the drone portion. The ``Truck'' column details the carbon emissions associated with truck transportation. The last two columns show the carbon emission deviations related to the Truck versus both the Bus-drone and the Drone. 

On one hand, the carbon emissions of trucks are lower than those of the bus-drone system in terms of Dev$_{BT}$. However, the bus line is predetermined, and the emissions from buses are inevitable, even when not carrying parcels. In contrast, the emissions from trucks are a result of the additional delivery process, which makes their impact more pronounced. As the problem size increases, Dev$_{BT}$ decreases, highlighting the emerging benefits of the bus-drone system. On the other hand, it is not surprising that the carbon emissions emitted by drones are significantly lower than those from trucks. On average, the carbon emissions of drones are less than 57.6 kg of CO$_2$. With the increase in the number of customers, Dev$_{DT}$ increasingly demonstrates the environmental friendliness of drones. These findings underscore the substantial environmental advantages of the integrated bus-drone system.

\subsubsection{Analysis of the Cost Parameters:}
\begin{figure}[htbp]
	\centering
	\includegraphics[width = 0.55 \textwidth]{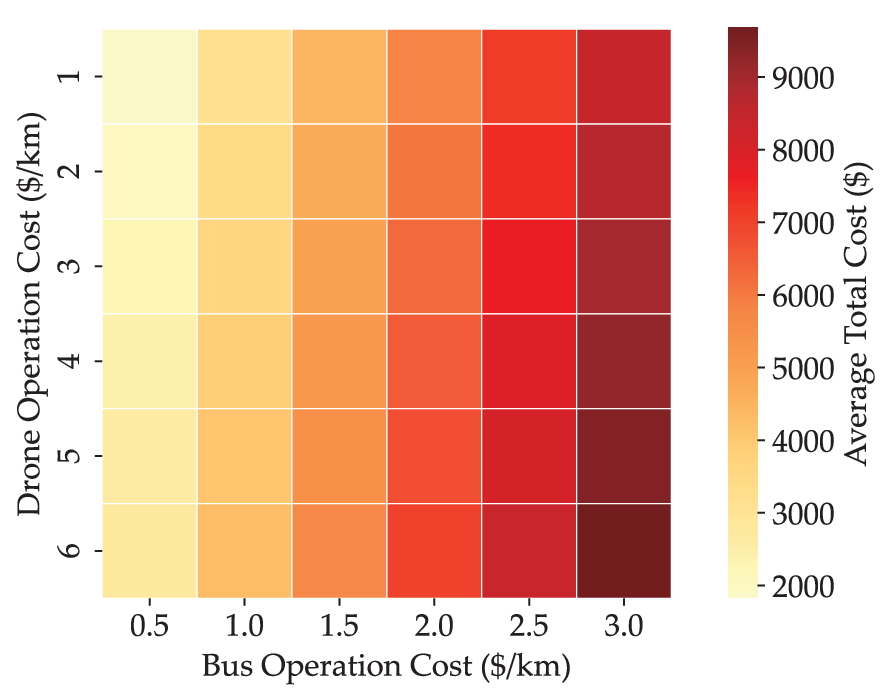}
	\caption{Total Cost as a Function of Bus and Drone Operation Costs.}	\label{figAnaCost}	
\end{figure}

The bus and drone operation costs ($c^1_{bs}$ and $c^2_{si}$) are pivotal components of the total cost in \prob. We use all instances in Customer27 as a baseline to investigate the sensitivity of the total cost to different combinations of bus and drone operation costs. Such an analysis can inform tailored cost management strategies, ultimately enhancing the system's overall operational efficiency. The experiment systematically varied the original bus and drone operation costs by a range of multiplicative factors from 0.5 to 3.0, examining a total of 36 different combinations. The average total cost across the 10 instances in Customer27 for each combination is depicted in each cell in Figure \ref{figAnaCost}. A darker color indicates a higher cost.

The heatmap shows that total costs are more sensitive to fluctuations in bus operation costs than in drone operation costs. In bus transportation, the parcel weight exacerbates the increase in bus operation costs during transportation, underscoring the need for efficient parcel management. Conversely, across different levels of bus operation costs, the total cost remains stable with various drone operation costs. This indicates the profitability potential of last-mile delivery by drones. The use of buses to expand drone service coverage enhances the profitability of drone services. Nonetheless, if bus operation costs are high, the logistics company should reconsider reliance on public transport, as the potential revenue may not offset these elevated operation costs. When both operation costs are under approximately \$2/km, the total cost increases at a moderate rate. This indicates that the system can operate with higher profit margins within this threshold, emphasizing the need for careful monitoring of the cost-effectiveness of integrating public transport with drone delivery services.



Apart from the bus and drone operation costs, there are two additional costs, namely the holding cost of a parcel and the fixed cost of a drone ($f^H$ and $f^F$). The sum of these two costs equates to the cost of the second-stage problem (BSP) and can reflect the performance of drones. Specifically, the profitability of drone delivery services depends on the maximum operation duration $\Delta$, as extended operation times can reduce the number of drones required. Furthermore, profits are affected by the holding cost $f^H$ of parcels in lockers since a higher $f^H$ may necessitate more drones to decrease parcel holding times. However, this can also be offset by improving the drone's velocity. Therefore, the next experiments examine the sensitivity of these two parameters to the costs of BSP. 

\begin{figure}[htbp]
	\begin{minipage}[t]{0.5\linewidth}
		\centering
		\includegraphics[width = 3.2in]{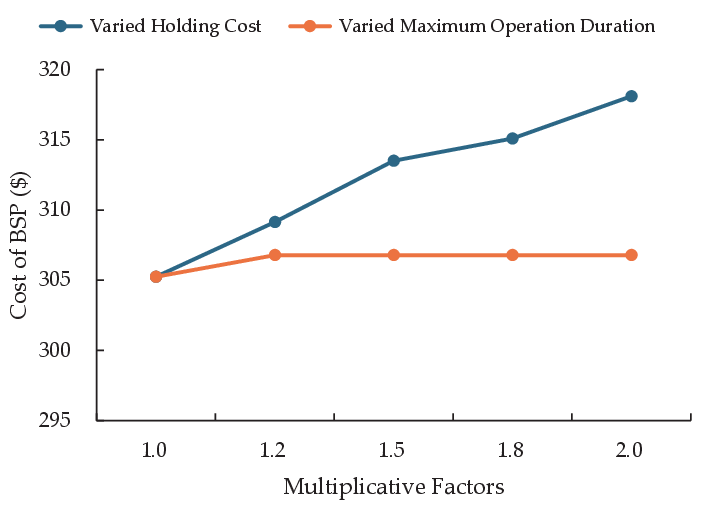}
		\centerline{(a) Cost of BSP as a Function of Locker and Drone.}
	\end{minipage}
	\begin{minipage}[t]{0.5\linewidth}
		\centering
		\includegraphics[width = 3.2in]{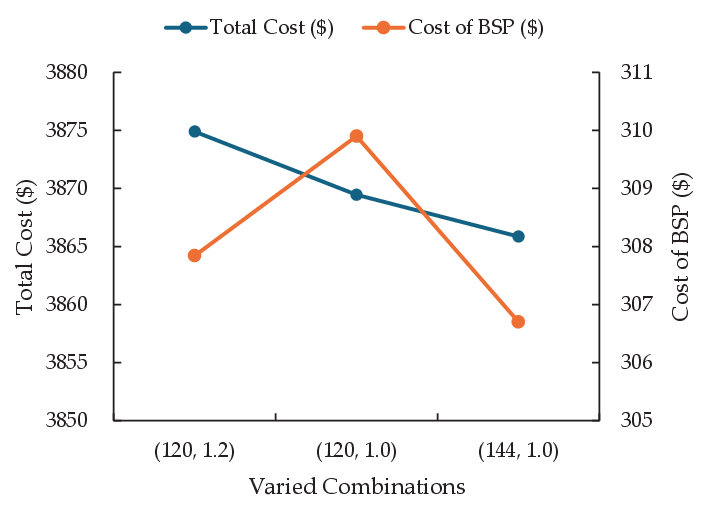}
		\centerline{(b) Cost of Instance Customer36-0.}	
	\end{minipage}
	\caption{Cost and a Relative Instance} \label{figAnaLD}	
\end{figure}

The experiments are conducted on the 10 instances in Customer36, chosen for its moderate customer demand size, with one of $\Delta$ or $f^H$ fixed while the other was varied using the multiplicative factors $\{1.0, 1.2, 1.5, 1.8, 2.0\}$ on their base values, respectively. The average cost across these 10 instances for each parameter combination is depicted in Figure \ref{figAnaLD}.(a). It is observed that an increase in the holding cost raises the cost of BSP, whereas the maximum operation duration has a small impact on the cost. This contrast highlights the significance of enhancing the efficiency of drone delivery, especially under high demand, which can escalate holding costs, thus increasing the financial incentive to reduce holding times.

Generally, the total cost increases with a decrease in $\Delta$ or an increase in $f^H$, as these changes can limit the feasible solutions. However, as shown in Figure \ref{figAnaLD}.(a), the fluctuations in the cost of BSP warrant a closer look at the instance level. Figure \ref{figAnaLD}.(b) takes the first instance in Customer36 for a further detailed analysis, with the x-axis displaying the varied combinations of $(\Delta, f^H)$ values. The results indicate that fluctuations in the total cost coincide with changes in the parameters, while an increase in the total cost does not necessarily result in a higher cost of BSP. This is because parcel assignments in the first stage can also influence the optimal solution, which once again underscores the benefits of joint optimization.

\subsubsection{Analysis of the Parcel Locker Configuration:}

This section investigates the impact of the parcel locker configuration on the performance of the bus-drone system. Installing lockers at every traversed bus stop is critical to the success of a multimodal freight transport system, yet it is impractical to do so due to the substantial infrastructure cost. Furthermore, the requirement for bus drivers to unload parcels at multiple stops, even during off-peak hours, extends the dwell times. To analyze the convenience of freight storage and passenger satisfaction, we consider a scenario with 40 customers randomly distributed across the demand region and examine three cases, varying the number of served customers to 20, 30, and 40, and adjusting the number of lockers to 5, 6, 7, and 8. 

\begin{figure}[htbp]
\centering
	\includegraphics[width = 0.5 \textwidth]{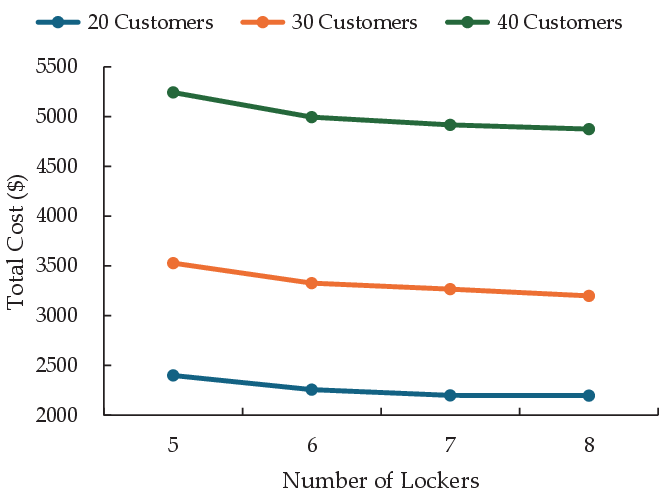}
	\caption{Analysis of the Number of Lockers.}	\label{figBusStop}	
\end{figure}

The results are presented in Figure \ref{figBusStop}. The figure indicates that the increased number of lockers reduces the total cost of the bus-drone system, as the denser locker network facilitates more convenient parcel delivery for drones. However, the trend of cost reduction plateaus once the locker density reaches a threshold where the drone coverage is broad enough for efficient service. Beyond that point, additional lockers offer diminishing returns in cost savings unless the number of customers increases. In this example, 7 lockers were found to be a suitable choice.

\section{Conclusions} \label{secConclusion}
This paper introduces a freight multimodal transport problem that integrates buses and drones, where parcels are transported from the bus terminal to parcel lockers at bus stops, and ultimately delivered to customers via drones. We optimize this integrated system by formulating a compact MILP model to minimize the total operation costs. To address large-scale instances, we propose an IP formulation with exponentially many variables and develop an exact algorithm that integrates column generation and Benders decomposition within a branch-and-bound framework. To boost algorithmic performance, several enhancements have also been introduced, including warm-start strategies, bounding-and-fixing strategies, valid inequalities, and primal heuristics. Numerical experiments are conducted using instances based on real-world bus data. Our results substantiate the effectiveness of our models and algorithms.

Valuable managerial insights are offered for the bus-drone system, gleaned from sensitivity analyses. First, both the exact and heuristic versions of the BPBC algorithm highlight the advantages of joint optimization over sequential optimization. Additionally, we discuss the environmental advantages of the bus-drone system, suggesting that it can significantly reduce carbon emissions, particularly as the problem size increases. We also look into the impact of the cost parameters and show that the bus operation cost and the parcel holding cost have a greater influence. Finally, the number of parcel lockers has a marginally diminishing effect on bus-drone system profitability. 


These findings pave the way for future research in the following directions. First, integrating the capacity of parcel lockers into our model, which was previously overlooked in favor of the holding cost, would create a more holistic bus-locker-drone system, optimizing the overall logistics ecosystem. Moreover, adapting the BPBC algorithm to handle uncertainties in customer demand presents an intriguing challenge, potentially making the algorithm more robust and adaptable to real-world fluctuations and unpredictability in logistics.


{
\SingleSpacedXI
\bibliographystyle{apalike}
\bibliography{reference}

\begin{thebibliography}{}

\bibitem[Agatz et~al., 2018]{Agatz2018}
Agatz, N., Bouman, P., and Schmidt, M. (2018).
\newblock Optimization approaches for the traveling salesman problem with
  drone.
\newblock {\em Transportation Science}, 52:965--981.

\bibitem[Amazon, 2024]{amazonair}
Amazon (2024).
\newblock Amazon {P}rime {A}ir.
\newblock \url{https://www.aboutamazon.com/news/tag/prime-air}.

\bibitem[Azcuy et~al., 2021]{Azcuy2021}
Azcuy, I., Agatz, N., and Giesen, R. (2021).
\newblock Designing integrated urban delivery systems using public transport.
\newblock {\em Transportation Research Part E}, 156:102525.

\bibitem[Barnhart et~al., 1998]{barnhart1998branch}
Barnhart, C., Johnson, E.~L., Nemhauser, G.~L., Savelsbergh, M.~W., and Vance,
  P.~H. (1998).
\newblock Branch-and-price: Column generation for solving huge integer
  programs.
\newblock {\em Operations Research}, 46:316--329.

\bibitem[Carlsson and Song, 2018]{Carlsson2018}
Carlsson, J.~G. and Song, S. (2018).
\newblock Coordinated logistics with a truck and a drone.
\newblock {\em Management Science}, 64:3971--4470.

\bibitem[{Chamber of Progress}, 2021]{chamber}
{Chamber of Progress} (2021).
\newblock Commercial delivery drones announced for {D}allas-{F}ort {W}orth
  area.
\newblock
  \url{https://progresschamber.org/commercial-delivery-drones-announced-for-dallas-fort-worth/}.

\bibitem[Cheng et~al., 2024]{Cheng2024}
Cheng, C., Adulyasak, Y., and Rousseau, L.-M. (2024).
\newblock Robust drone delivery with weather information.
\newblock {\em Manufacturing \& Service Operations Management}, 26:1402--1421.

\bibitem[Cheng et~al., 2023]{Cheng2023}
Cheng, R., Jiang, Y., Nielsen, O.~A., and Pisinger, D. (2023).
\newblock An adaptive large neighborhood search metaheuristic for a passenger
  and parcel share-a-ride problem with drones.
\newblock {\em Transportation Research Part C}, 153:104203.

\bibitem[Choudhury et~al., 2021]{Choudhury2021}
Choudhury, S., Solovey, K., Kochenderfer, M.~J., and Pavone, M. (2021).
\newblock Efficient large-scale multi-drone delivery using transit networks.
\newblock {\em Journal of Artificial Intelligence Research}, 70:757--788.

\bibitem[Costa et~al., 2019]{Costa2019}
Costa, L., Contardo, C., and Desaulniers, G. (2019).
\newblock Exact branch-price-and-cut algorithms for vehicle routing.
\newblock {\em Transportation Science}, 53:946--985.

\bibitem[Daryalal et~al., 2023]{Daryalal2023}
Daryalal, M., Pouya, H., and DeSantis, M.~A. (2023).
\newblock Network migration roblem: A hybrid logic-based benders decomposition
  approach.
\newblock {\em INFORMS Journal of Computing}, 35:593--613.

\bibitem[Desaulniers et~al., 2005]{Desaulniers2005}
Desaulniers, G., Desrosiers, J., and Solomon, M.~M. (2005).
\newblock {\em Column Generation}.
\newblock Springer, New York, USA.

\bibitem[Di et~al., 2022]{Di2022}
Di, Z., Yang, L., Shi, J., Zhou, H., Yang, K., and Gao, Z. (2022).
\newblock Joint optimization of carriage arrangement and flow control in a
  metro-based underground logistics system.
\newblock {\em Transportation Research Part B}, 159:1--23.

\bibitem[Dies, 2022]{Dies2022}
Dies, J. (2022).
\newblock Parcel shipping index 2022 featuring 2021 data.
\newblock Technical report, Pitney Bowes.

\bibitem[Donne et~al., 2023]{Donne2023}
Donne, D.~D., Alfandari, L., Archetti, C., and Ljubić, I. (2023).
\newblock Freight-on-transit for urban last-mile deliveries: A strategic
  planning approach.
\newblock {\em Transportation Research Part B}, 169:53--81.

\bibitem[dos Santos et~al., 2022]{dosSantos2022}
dos Santos, A.~G., Viana, A., and Pedroso, J.~P. (2022).
\newblock 2-echelon lastmile delivery with lockers and occasional couriers.
\newblock {\em Transportation Research Part E}, 162:102714.

\bibitem[Dror, 1994]{dror1994note}
Dror, M. (1994).
\newblock Note on the complexity of the shortest path models for column
  generation in {VRPTW}.
\newblock {\em Operations Research}, 42:977--978.

\bibitem[Eggleston, 2006]{Eggleston2006}
Eggleston, S. (2006).
\newblock Estimation of emissions from {CO}2 capture and storage: {T}he 2006
  {IPCC} guidelines for national greenhouse gas inventories.
\newblock In {\em Presentation at the UNFCCC Workshop on Carbon Dioxide Capture
  and Storage}, volume~20.

\bibitem[Florio et~al., 2020]{florio2020new}
Florio, A.~M., Hartl, R.~F., and Minner, S. (2020).
\newblock New exact algorithm for the vehicle routing problem with stochastic
  demands.
\newblock {\em Transportation Science}, 54(4):1073--1090.

\bibitem[Ghilas et~al., 2018]{Ghilas2018}
Ghilas, V., Cordeau, J.-F., Demir, E., and Woensel, T.~V. (2018).
\newblock Branch-and-price for the pickup and delivery problem with time
  windows and scheduled lines.
\newblock {\em Transportation Science}, 52:1191--1210.

\bibitem[Goodchild and Toy, 2018]{Goodchild2018}
Goodchild, A. and Toy, J. (2018).
\newblock Delivery by drone: An evaluation of unmanned aerial vehicle
  technology in reducing {CO}2 emissions in the delivery service industry.
\newblock {\em Transportation Research Part D}, 61:58--67.

\bibitem[Gross, 2013]{Gross2013}
Gross, D. (2013).
\newblock Amazon’s drone delivery: How would it work?
\newblock
  \url{https://edition.cnn.com/2013/12/02/tech/innovation/amazon-drones-questions/}.

\bibitem[He et~al., 2022]{He2022}
He, L., Liu, S., and Shen, Z.-J.~M. (2022).
\newblock Smart urban transport and logistics: A business analytics
  perspective.
\newblock {\em Production and Operations Management}, 31:3771--3787.

\bibitem[Karsten et~al., 2018]{Karsten2018}
Karsten, C.~V., Ropke, S., and Pisinger, D. (2018).
\newblock Simultaneous optimization of container ship sailing speed and
  container routing with transit time restrictions.
\newblock {\em Transportation Science}, 52:769--787.

\bibitem[{Kunming Bus}, 2022]{Kunming2022}
{Kunming Bus} (2022).
\newblock Kunming public transport launches urban freight ``public transit''
  logistics cooperation project (in {C}hinese).
\newblock {\em Urban Public Transport}, 4:58.

\bibitem[Kızıl and Yıldız, 2023]{Kizil2023}
Kızıl, K.~U. and Yıldız, B. (2023).
\newblock Public transport-based crowd-shipping with backup transfers.
\newblock {\em Transportation Science}, 57:174--196.

\bibitem[Li et~al., 2021]{Li2021}
Li, Z., Shalaby, A., Roorda, M.~J., and Mao, B. (2021).
\newblock Urban rail service design for collaborative passenger and freight
  transport.
\newblock {\em Transportation Research Part E}, 147:102205.

\bibitem[Madani et~al., 2024]{Madani2024}
Madani, B., Ndiaye, M., and Salhi, S. (2024).
\newblock Hybrid truck-drone delivery system with multi-visits and multi-launch
  and retrieval locations: Mathematical model and adaptive variable
  neighborhood search with neighborhood categorization.
\newblock {\em European Journal of Operational Research}, 316:100--125.

\bibitem[Masson et~al., 2017]{Masson2017}
Masson, R., Trentini, A., Lehu{\'e}d{\'e}, F., Malh{\'e}n{\'e}, N., P{\'e}ton,
  O., and Tlahig, H. (2017).
\newblock Optimization of a city logistics transportation system with mixed
  passengers and goods.
\newblock {\em EURO Journal on Transportation and Logistics}, 6:81--109.

\bibitem[Moadab et~al., 2022]{Moadab2022}
Moadab, A., Farajzadeh, F., and Valilai, O.~F. (2022).
\newblock Drone routing problem model for last‑mile delivery using the public
  transportation capacity as moving charging stations.
\newblock {\em Scientific Reports}, 12:6361.

\bibitem[Murray and Chu, 2015]{Murray2015}
Murray, C.~C. and Chu, A.~G. (2015).
\newblock The flying sidekick traveling salesman problem: Optimization of
  drone-assisted parcel delivery.
\newblock {\em Transportation Research Part C}, 54:86--109.

\bibitem[Naderi et~al., 2021]{Naderi2021}
Naderi, B., Roshanaei, V., Begen, M.~A., Aleman, D.~M., and Urbach, D.~R.
  (2021).
\newblock Increased surgical capacity without additional resources: Generalized
  operating room planning and scheduling.
\newblock {\em Production and Operations Management}, 30:2608--2635.

\bibitem[Perera et~al., 2020]{Perera2020}
Perera, S., Dawande, M., Janakiraman, G., and Mookerjee, V. (2020).
\newblock Retail deliveries by drones: How will logistics networks change?
\newblock {\em Production and Operations Management}, 29:2019--2034.

\bibitem[Pessoa et~al., 2020]{pessoa2020generic}
Pessoa, A., Sadykov, R., Uchoa, E., and Vanderbeck, F. (2020).
\newblock A generic exact solver for vehicle routing and related problems.
\newblock {\em Mathematical Programming}, 183:483--523.

\bibitem[Poikonen et~al., 2019]{Poikonen2019}
Poikonen, S., Golden, B.~L., and Wasil, E.~A. (2019).
\newblock A branch-and-bound approach to the traveling salesman problem with a
  drone.
\newblock {\em INFORMS Journal on Computing}, 31:335--346.

\bibitem[Pugliese et~al., 2021]{Pugliese2021}
Pugliese, L. D.~P., Macrina, G., and Guerriero, F. (2021).
\newblock Trucks and drones cooperation in the last-mile delivery process.
\newblock {\em Networks}, 78:371--399.

\bibitem[Restrepo et~al., 2018]{Restrepo2018}
Restrepo, M.~I., Gendron, B., and Rousseau, L.-M. (2018).
\newblock Combining benders decomposition and column generation for
  multi-activity tour scheduling.
\newblock {\em Computers and Operations Research}, 93:151--165.

\bibitem[Rostami et~al., 2021]{rostami2021branch}
Rostami, B., Desaulniers, G., Errico, F., and Lodi, A. (2021).
\newblock Branch-price-and-cut algorithms for the vehicle routing problem with
  stochastic and correlated travel times.
\newblock {\em Operations Research}, 69:436--455.

\bibitem[Sacramento et~al., 2019]{Sacramento2019}
Sacramento, D., Pisinger, D., and R{\o}pke, S. (2019).
\newblock An adaptive large neighborhood search metaheuristic for the vehicle
  routing problem with drones.
\newblock {\em Transportation Research Part C}, 102:289--315.

\bibitem[Sahli et~al., 2022]{Sahli2022}
Sahli, A., Behiri, W., Belmokhtar-Berraf, S., and Chu, C. (2022).
\newblock An effective and robust genetic algorithm for urban freight transport
  scheduling using passenger rail network.
\newblock {\em Computers \& Industrial Engineering}, 173:108645.

\bibitem[Schmidt et~al., 2024]{schmidt2024using}
Schmidt, J., Tilk, C., and Irnich, S. (2024).
\newblock Using public transport in a 2-echelon last-mile delivery network.
\newblock {\em European Journal of Operational Research}, 317:827--840.

\bibitem[Stodola and Kutěj, 2024]{Stodola2024}
Stodola, P. and Kutěj, L. (2024).
\newblock Multi-depot vehicle routing problem with drones: Mathematical
  formulation, solution algorithm and experiments.
\newblock {\em Expert Systems with Applications}, 241:122483.

\bibitem[Sun et~al., 2023]{Sun2023}
Sun, X., Hu, W., Xue, X., and Dong, J. (2023).
\newblock Multi-objective optimization model for planning metro-based
  underground logistics system network: Nanjing case study.
\newblock {\em Journal of Industrial and Management Optimization}, 19:170--196.

\bibitem[{Walmart}, 2024]{Walmart}
{Walmart} (2024).
\newblock Sky high ambitions: Walmart to make largest drone delivery expansion
  of any u.s. retailer.
\newblock
  \url{https://corporate.walmart.com/news/2024/01/09/sky-high-ambitions-walmart-to-make-largest-drone-delivery-expansion-of-any-us-retailer}.

\bibitem[Wang and Sheu, 2019]{Wang2019}
Wang, Z. and Sheu, J.-B. (2019).
\newblock Vehicle routing problem with drones.
\newblock {\em Transportation Research Part B}, 122:350--364.

\bibitem[Wei et~al., 2020]{Wei2020}
Wei, L., Luo, Z., Baldacci, R., and Lim, A. (2020).
\newblock A new branch-and-price-and-cut algorithm for one-dimensional
  bin-packing problems.
\newblock {\em INFORMS Journal on Computing}, 32:428--443.

\bibitem[Wu et~al., 2023]{Wu2023}
Wu, Y., Wang, S., Zhen, L., Laporte, G., Tan, Z., and Wang, K. (2023).
\newblock How to operate ship fleets under uncertainty.
\newblock {\em Production and Operations Management}, 32:3043--3061.

\bibitem[Yang et~al., 2023]{Yang2023}
Yang, X., Wu, W., and Huang, G.~Q. (2023).
\newblock A crowdsourced co-modality transportation system integrating
  passenger and freight.
\newblock {\em Advanced Engineering Informatics}, 58:102142.

\bibitem[Yu et~al., 2022]{Yu2022}
Yu, Q., Cheng, C., and Zhu, N. (2022).
\newblock Robust team orienteering problem with decreasing profits.
\newblock {\em INFORMS Journal on Computing}, 34:3215--3233.

\bibitem[Zhu et~al., 2024]{Zhu2024}
Zhu, W., Hu, X., Pei, J., and Pardalos, P.~M. (2024).
\newblock Minimizing the total travel distance for the locker-based drone
  delivery: A branch-and-cut-based method.
\newblock {\em Transportation Research Part B}, 184:102950.

\bibitem[Zou et~al., 2024]{Zou2023}
Zou, B., Wu, S., Gong, Y., Yuan, Z., and Shi, Y. (2024).
\newblock Delivery network design of a locker-drone delivery system.
\newblock {\em International Journal of Production Research}, 62:4097--4121.

\end{thebibliography}
}
\ECSwitch

\vspace{-.8cm}
\ECHead{Electronic Companion}

\vspace{-.4cm}
{\TableSpaced
	\begin{longtable}[htbp]{lp{13cm}}
		\caption{List of the Key Notations and Decision Variables.} \vspace{-.1cm}\label{tabNotations} \\
		\toprule
		Notation 		& Definition\\ 	
		\midrule
		\textbf{Sets and indexes}\\
		$G = (\mcV, \mcA)$	& Graph representing the optimization problem\\
		$\mcV$				& Set of vertices on the graph\\
		$\mcA$				& Set of arcs on the graph\\
		$\{o\}$			& Terminal\\
		$\mcS$				& Set of bus stops or parcel lockers, $s \in \mcS$\\
		$\mcC$				& Set of customers, $i \in \mcC$\\
		$\mcB$				& Set of buses, $b \in \mcB$\\
		$\mcD$				& Set of drones, $d \in \mcD$\\
		$\mcC_s$			& Subset of customers reachable by drones from locker $s$\\
		$\mcB_s(i)$		& Subset of buses capable of delivering demand $i$ to stop $s$\\  
		$\mcD_s$			& Subset of drones that can perform deliveries from locker $s$\\
		$\mcT$&Set of all feasible assignments \\
		$\mcT_{bs}$&Set of assignments associated with bus $b$ at stop $s$ \\
		$\mcU$& Set of feasible flights\\
		$\mcU_s$&Set of feasible flights associated with locker $s$\\\midrule
		\textbf{Parameters}\\
		$q_i$ 			& The amount of parcel boxes of customer demand $i$, note that $q_o = 0$\\
		$Q^B$			& Bus capacity for parcel transportation\\
		$Q^S$			& The maximum number of boxes that can be unloaded at the stop\\
		$l_i$ 			& Service deadline of customer $i$, note that $[0, l_o]$ represents the planning horizon\\
		$e_{bs}$		& Arrival time of bus $b$ at stop $s$\\
		$\tau_{ij}$		& Travel time of drone arc $(i,j) \in \mcA$\\
		$\tau^B$		& Fixed time for a bus to unload boxes at the stop\\
		$\tau^S$		& Fixed time for a drone to load boxes and replace a battery in the locker\\
		$\tau^D$		& Fixed time for a drone to deliver boxes to a customer\\
		$\delta_{si}$	& Total operation duration for a round trip of a drone from stop $s$ to customer $i$\\
		$\Delta$		& Maximum operation duration of each drone\\
		$c_{bs}^1$		& Operation cost per unit of freight on bus $b$ to stop $s$\\
		$c_{si}^2$ 		& Operation cost for a round trip of a drone from stop $s$ to customer $i$\\
		$f^H$			& Holding cost per unit of time and per unit of freight at a locker\\
		$f^F$			& Daily maintenance and purchase cost (fixed cost) of a drone\\
		$\alpha_{it}$ & Binary coefficient indicating whether customer $i$ is assigned to assignment $t$\\
		$\beta_t$ & The total load in assignment $t$\\
		$\gamma_{ibu}$& Binary coefficient indicating whether flight $u$ serves customer $i$ unloaded by bus $b$\\\midrule
		\textbf{Decision variables} & \textbf{in MILP1}\\
		$x_{sib}$ 		& One, if bus $b$ delivers demand $i$ to stop $s$, and zero otherwise\\
		$y_{ijsd}$		& One, if drone $d$ associated with locker $s$ serves (1) customer $i$ before customer $j$, (2) customer $j$ as the first customer from locker $i$, or (3) customer $i$ as the last customer from locker $j$, and zero otherwise\\
		$z_{sd}$		& One, if drone $d$ is launched from locker $s$, and zero otherwise\\
		$h_i$			& Holding time of demand $i$\\
		$w_i$ 			& Service start time of the drone at customer $i$\\\midrule
		\textbf{Decision variables} & \textbf{in TCF}\\
		$\theta_t$ &One, if assignment $t$ is selected, and zero otherwise \\
		$\zeta_u$& One, if flight $u$ is selected, and zero otherwise\\
		\bottomrule
	\end{longtable}
}


\section{Proofs of Statements} \label{appProve}

\subsection{Proof of Theorem \ref{proDBSP}}
\proof{Proof:} Given that $c_u^2 \geq 0$ in constraints (\ref{DBSP:visit}), the null vector $\mathbold{0}$ for $\mathbold{\omega}$ satisfies DBSP$_s$($\boldsymbol{\theta}$) and hence, the feasible domain of DBSP$_s$($\boldsymbol{\theta}$) is not empty. By strong duality, BSP$_s$($\boldsymbol{\theta}$) is either bounded and feasible or infeasible. Given that we assume we have sufficient drones and the definition of $\mcB_s(i)$, we can construct a feasible solution that each flight only serves a customer that satisfies $\alpha_{it}\theta_t > 0$. Therefore, BSP$_s$($\boldsymbol{\theta}$) is bounded and feasible, and so is DBSP$_s$($\boldsymbol{\theta}$).   $\hfill\Box$
\endproof

\subsection{Proof of Proposition \ref{proBound}}
\proof{Proof:} The proof focuses on the origin BMP without (\ref{lowerbound}) because the same argument can be extended to the BMP with more constraints. Recall that $\mathbold{\mu} \geq 0$, $\mathbold{\pi} \leq 0$, $\mathbold{\lambda} \leq 0$, and $\mathbold{\rho} \geq 0$ are the dual variables associated with constraints (\ref{TCF:visit}) -- (\ref{TCF:Capacity}), and (\ref{BMP:ClassicCut}), respectively. Further, $\Gamma$ is the optimal value of the current BMP with the $\theta$ variables and constraints~(\ref{BMP:ClassicCut}) generated so far, and $v_{bs}$ is the reduced cost of an optimal column found in MPS$_{bs}$. By strong duality and definition, we have 
	\begin{equation} \label{eq1}
		\Gamma = \sum_{i \in \mcC} \mu_i + \sum_{b \in \mcB}\sum_{s \in \mcS} \pi_{bs} + \sum_{b \in \mcB} Q^B\lambda_b
	\end{equation}
	\begin{equation}
		v_{bs} = \min_{t \in \mcT_{bs}} \{c_1^t  - \sum_{i \in \mcC}\alpha_{it}\mu_i - \pi_{bs} - \beta_t\lambda_b+ \sum_{\mathbold{\omega}\in \Omega_s}\sum_{i \in \mcC_s} \rho_{\mathbold{\omega}}\alpha_{it}\omega_{ib}\}
	\end{equation}
	By the feasibility of the dual of BMP, the dual constraints regarding $\varphi_s$ for all $s$ in $\mcS$ are
	\begin{equation}
		\sum_{\mathbold{\omega} \in \Omega_s} \rho_{\mathbold{\omega}} \leq 1, ~\forall~ s \in \mcS
	\end{equation}
	Let $Z^*$ and $(\mathbold{\theta}^*, \mathbold{\varphi}^*)$ be the optimal value and optimal solutions of the full BMP with all $\theta$ variables and constraints~(\ref{BMP:ClassicCut}), respectively. By the feasibility of the solution of BMP, we have 
	\begin{equation}	\label{eq3}
		\sum_{s \in \mcS}\sum_{b \in \mcB}\sum_{t \in \mcT_{bs}} \alpha_{it}\theta_t^* \geq 1, ~\forall~ i \in \mcC
	\end{equation}
	\begin{equation}	
		\sum_{t \in \mcT_{bs}} \theta_t^* \leq 1, ~\forall~ b \in \mcB, s \in \mcS
	\end{equation}
	\begin{equation}	
		\sum_{s \in \mcS}\sum_{t \in \mcT_{bs}} \beta_t\theta_t^* \leq Q^B, ~\forall~ b \in \mcB
	\end{equation}
	\begin{equation} \label{eq2}	
		\varphi_s^* \geq \sum_{i \in \mcC_s}\sum_{b \in \mcB_s(i)}\sum_{t \in \mcT_{bs}} \alpha_{it}\omega_{ib}\theta_t^*, ~\forall~ s \in \mcS, \boldsymbol{\omega} \in \Omega_s
	\end{equation}
	Due to (\ref{eq1}) -- (\ref{eq2}), we can deduce that
	\begin{align*}
		Z^* - \Gamma &= \sum_{t \in \mcT} c_1^t\theta_t^* + \sum_{s \in \mcS} \varphi_s^* - (\sum_{i \in \mcC} \mu_i + \sum_{s \in \mcS}\sum_{b \in \mcB} \pi_{bs} + \sum_{b \in \mcB} Q^B\lambda_b)\\
		&\geq \sum_{t \in \mcT} c_1^t\theta_t^* 
		+ \sum_{s \in \mcS}\sum_{\mathbold{\omega} \in \Omega_s}\sum_{i \in \mcC_s}\sum_{b \in \mcB_s(i)}\sum_{t \in \mcT_{bs}} \rho_{\mathbold{\omega}}\alpha_{it}\omega_{ib}\theta_t^* 
		- \sum_{i \in \mcC}\sum_{s \in \mcS}\sum_{b \in \mcB}\sum_{t \in \mcT_{bs}} \alpha_{it}\theta_t^*\mu_i\\ 
		&\quad  - \sum_{t \in \mcT_{bs}} \theta_t^*\pi_{bs} 
		- \sum_{s \in \mcS}\sum_{t \in \mcT_{bs}} \beta_t\theta_t^*\lambda_b\\
		&=\sum_{s \in \mcS}\sum_{b \in \mcB}\sum_{t \in \mcT_{bs}} \theta_t^*(c_1^t + \sum_{\mathbold{\omega} \in \Omega_s}\sum_{i \in \mcC_s} \rho_{\mathbold{\omega}}\alpha_{it}\omega_{ib} - \sum_{i \in \mcC}\alpha_{it}\mu_i - \pi_{bs} - \beta_t\lambda_b)\\
		&\geq \sum_{s \in \mcS}\sum_{b \in \mcB}\sum_{t \in \mcT_{bs}} \theta_t^* v_{bs}\\
		&\geq \sum_{s \in \mcS}\sum_{b \in \mcB: v_{bs} < 0} v_{bs},
	\end{align*}
	which completes the proof. The first inequality is obtained by following \eqref{eq3}-\eqref{eq2} directly. The second inequality is due to the definition of $v_{bs}$. The last inequality follows due to the fact that we focus on $v_{bs} < 0$ and $0\leq\sum_{t\in\mcT_{bs}} \theta_t^*\leq 1$. for all $b$ in $\mcB$ and $s$ in $\mcS$. $\hfill\Box$
\endproof



\subsection{Proof of Proposition \ref{proLowerbound}}
\proof{Proof:} The validity of Proposition \ref{proLowerbound} can be proved by the following two cases.

\textbf{Case 1:} when no parcel is assigned to stop $s$. In this case, the right-hand side of inequality (\ref{lowerbound}) equals zero. Hence, the inequality is satisfied, given that $\varphi_s \geq 0$.
	
\textbf{Case 2:} when at least one parcel is assigned to stop $s$. In this case, $\varphi_s$ includes the holding cost of parcels and the fixed cost of drones. Since the holding cost is also related to the drone scheduling that is solved in BSP, we use the fixed cost as a lower bound for $\varphi_s$. Note that the total operation duration is at least the sum of the operation duration of the customers assigned to it. Thus, we can calculate the minimum number of drones used by assigning it evenly to each drone, which corresponds to the fraction on the right-hand side. 
	
To conclude, the inequalities (\ref{lowerbound}) are valid.    $\hfill\Box$
\endproof

\section{Details on Column Generation} \label{detail-secCG}
We elaborate on the column generation approaches by first introducing the pricing subproblems of both BMP and BSP, followed by the labeling algorithm for their solutions.

\subsection{Pricing Problem of BMP}
The pricing problem of BMP, denoted as MPS, aims to search for the feasible column with the minimum reduced cost for each bus at each stop. The MPS consists of $|B||S|$ distinct subproblems, each corresponding to a bus $b$ that departs from the terminal on its distinct schedule and arrives at a stop $s$, denoted as MPS$_{bs}$. 

Let $\mathbold{\mu} \geq 0$, $\mathbold{\pi} \leq 0$, $\mathbold{\lambda} \leq 0$, and $\mathbold{\rho} \geq 0$ be the dual variables associated with constraints (\ref{TCF:visit}) -- (\ref{TCF:Capacity}), and (\ref{BMP:ClassicCut}), respectively. We define $\Theta_{sib}^1 = c^1_{bs} q_i + c^2_{si} - \mu_i - q_i \lambda_b + \sum_{\mathbold{\omega} \in \Omega_s} \omega_{ib} \rho_{\mathbold{\omega}}$ as the contribution to the reduced cost for assigning customer $i$ to bus $b$ at stop $s$. Each MPS$_{bs}$ is formulated as 
\begin{equation}
	\min_{t \in \mcT_{bs}}\bigg(- \pi_{bs} + \sum_{i \in \mcC_s} \alpha_{it}\Theta_{sib}^1\bigg)
\end{equation}
that identifies the optimal assignment $t \in \mcT_{bs}$ with the minimum reduced cost. We define MPS$_{bs}$ on a logical graph $G_{bs} = (\mcV_{bs}, \mcA_{bs})$. The vertex set $\mcV_{bs}$ comprises two dummy source and sink vertices $\{s, s^{\prime}\}$, and a set of customers $\mcN_t$ eligible for assignment to bus $b$ at stop $s$. A unique order is imposed on the elements of $\mcN_t$, where $i \prec j$ indicates that customer $i$ precedes customer $j$ in this order. The set of arcs $\mcA_{bs}$ consists of three types: arcs from the source to customers, arcs from customers to the sink, and arcs between two customers $(i,j)$ if and only if $i \prec j$ for symmetry breaking. We also need to consider the maximum number of boxes that can be unloaded at a bus stop ($Q^S$). By this way, MPS$_{bs}$ is a variant of the elementary shortest path problem with resource constraints (ESPPRC), which is NP-hard in the strong sense \citep{dror1994note}. 

\subsection{Pricing Problem of BSP}
After solving BMP, we obtain $\mathbold{\theta}$, which provides a set of assignments for each stop. The pricing problem of BSP is denoted as SPS. Recall that there are $|S|$ subproblems in SPS. Each pricing subproblem, denoted as SPS$_s$, aims to search for a feasible drone flight with the minimum reduced cost associated with locker $s$. 

Denote $e_i^b = e_{bs} + \tau^B$ as the release time for parcels $i$, indicating the earliest time at which the parcel becomes available for delivery and the holding cost incurs. We define SPS$_s$ on a space-time graph $G_s = (\mcV_s, \mcA_s)$, where $\mcV_s$ and $\mcA_s$ are vertex set and arc set, respectively. The vertex set $\mcV_s$ includes the source and the sink vertices $\{s, s^{\prime}\}$, and a set of the vertices $\mcN_s$ indicating the customers and their corresponding buses. That is, $\mcN_s = \{(i, b)| \sum_{t \in \mcT_{bs}} \alpha_{it}\theta_t > 0, i \in \mcC_s, b \in \mcB_s(i)\}$. The arc set $\mcA_s$ is defined as $\mcA_s = (\{s\} \times \mcN_s) \cup (\mcN_s \times \{s^{\prime}\}) \cup \mcE_s$, where $\mcE_s = \{[(i,b), (j, b^{\prime})] | \max\{e_i^b + \delta_{si}, e_j^{b^{\prime}}\} + \tau^S + \tau_{si} \leq l_j, (i,b), (j, b^{\prime}) \in \mcN_s, i \neq j\}$. Furthermore, the service time of customer $i$ is constrained by $l_i$, and the operation duration of a drone is limited by $\Delta$. Consequently, SPS$_s$ can also be modeled as a variant of the ESPPRC, aiming to identify the minimal-reduced-cost flight given by
\begin{equation}
	\min_{u \in \mcU_s}\bigg(f^F - \sum_{(i,b) \in \mcN_s} \gamma_{ibu} \Theta_{ib}^2\bigg)
\end{equation}
where $\Theta_{ib}^2 = f^H(h_i - e_i^b) - \omega_{ib}$ is the reduced cost for serving vertex $(i,b)$.

\subsection{Labeling Algorithm} \label{seclabeling}
Considering the complexity of the ESPPRC, a labeling algorithm is proposed to identify new columns with negative reduced costs for both MPS and SPS. This algorithm is a widely used dynamic programming approach to generate feasible complete paths from the source to the sink on the ESPPRC graph. It has shown remarkable efficacy in solving various pricing subproblems within the column generation method \citep{Wei2020, Yu2022}.

The algorithm initiates at source $s$ and iteratively extends partial paths to new vertices on the graph. Let $\mcR$ be the set of resources. Each partial path ending at vertex $i$ is represented by a \textit{label} $L_i = (\bar{c}_i, \{r_i\}_{r \in \mcR})$, which records the accumulated reduced cost $\bar{c}_i$ and the resource consumption $r_i$ for every $r \in \mcR$. For MPS$_{bs}$, resource consumption of $L_i$ includes customer demand $r_i^q$ and a binary visit-counter $r_i^\nu$ for each customer in $\mcN_t$. For SPS$_s$, the label's resource consumption includes service time $r_i^t$, operation duration $r_i^\delta$, and a binary visit-counter $r_i^\nu$ for each customer in $\mcN_s$. Note that resource consumption at a node is applied to the incoming arcs of that node. 

During the extension of label $L_i$ to create label $L_j$ over arc $(i,j)$, all attributes of $L_i$ are updated via \textit{resource extension functions} that accumulate the reduced cost and compute the resource consumption associated with the traversed arc. Specifically, in an MPS$_{bs}$, $L_j = (\bar{c}_i + \Theta_{sjb}^1, r_i^q + q_j, r_i^\nu \cup \{j\}) $. In an SPS$_s$, $L_j = (\bar{c}_i + \Theta_{jb}^2, \max\{r_i^t + \delta_{si}, e_j^{b'}\} + \tau^S + \tau_{si}, r_i^\delta + \delta_{sj}, r_i^\nu \cup \{j\})$. The updated consumption must satisfy the corresponding constraints of the respective pricing subproblems. In MPS$_{bs}$, customer demand is bounded by the stop capacity $Q^S$, while in SPS$_s$, service time and operation duration are capped by $l_j$ and $\Delta$, respectively. Both binary visit-counters ensure that the columns are elementary.

Once the partial path reaches the sink, a feasible column corresponding to the complete path is obtained. Although the algorithm can enumerate all labels, the process is expedited by applying the following \textit{dominance rules} to discard unpromising labels \citep{Desaulniers2005}:
\begin{definition}[Dominance rule 1]
	For two labels $L_1$ and $L_2$ ending at the same vertex, $L_1$ dominates $L_2$ if: (i) $\bar{c}_1 \leq \bar{c}_2$, and (ii) $r_1 \leq r_2, ~\forall~ r \in \mcR$.
\end{definition}

Dominance rules imply that any feasible extension of $L_2$ is also feasible for $L_1$, and the reduced cost of the complete path derived from extending $L_1$ is guaranteed to be no worse than that obtained by extending $L_2$. Finally, we add the identified assignments (resp. flights) as new columns to resolve BMP (resp. BSP). The algorithm stops when no further assignments (resp. flights) with negative reduced costs are identified.

\section{Additional Details about the Computational Experiments} \label{appDetailed}
\begin{table}[htbp]
	\centering	
	\TableSpaced
	\caption{Detailed Instance-Level Results of the Overall Performance.}
	\resizebox{\textwidth}{!}{
    \begin{tabular}{lllllllllllll}
    \toprule
    Instance &       & UB    &       & Basic &       & Basic-W &       & Basic-WB &       & Basic-WBI &       & BPBC (Basic-WBIH) \\
    \midrule
Customer18-0 &       & 2259.55  &       & 0.9   &       & 0.6   &       & 1.0   &       & 1.5   &       & 1.5  \\
    Customer18-1 &       & 1996.95  &       & 6.2   &       & 2.6   &       & 5.0   &       & 3.3   &       & 3.2  \\
    Customer18-2 &       & 2141.85  &       & 0.7   &       & 0.3   &       & 0.2   &       & 1.1   &       & 1.1  \\
    Customer18-3 &       & 2556.85  &       & 2.3   &       & 1.5   &       & 3.2   &       & 2.3   &       & 2.6  \\
    Customer18-4 &       & 2592.65  &       & 1.7   &       & 1.2   &       & 1.7   &       & 2.1   &       & 2.0  \\
    Customer18-5 &       & 1828.25  &       & 0.3   &       & 0.2   &       & 0.2   &       & 1.2   &       & 1.2  \\
    Customer18-6 &       & 2283.70  &       & 3.6   &       & 1.9   &       & 0.6   &       & 2.2   &       & 2.0  \\
    Customer18-7 &       & 2431.35  &       & 1.0   &       & 0.8   &       & 0.7   &       & 1.5   &       & 1.8  \\
    Customer18-8 &       & 2477.85  &       & 0.7   &       & 0.5   &       & 0.4   &       & 1.2   &       & 1.3  \\
    Customer18-9 &       & 2451.50  &       & 0.5   &       & 0.3   &       & 0.4   &       & 1.1   &       & 1.1  \\
Customer27-0 &       & 3840.35  &       & 166.5  &      & 141.0  &       & 119.7  &       & 166.5  &       & 108.4  \\
Customer27-1 &       & 3002.35  &       & 6.8   &       & 8.1   &       & 8.5   &       & 5.4   &       & 7.8  \\
Customer27-2 &       & 2844.10  &       & 27.4  &       & 28.4  &       & 27.8  &       & 23.8  &       & 21.5  \\
Customer27-3 &       & 3245.90  &       & 37.9  &       & 35.6  &       & 35.5  &       & 26.5  &       & 27.3  \\
Customer27-4 &       & 3329.25  &       & 3.9   &       & 3.8   &       & 3.8   &       & 7.4   &       & 7.4  \\
Customer27-5 &       & 4076.05  &       & 22.4  &       & 22.6  &       & 19.6  &       & 22.4  &       & 25.8  \\
Customer27-6 &       & 3019.45  &       & 46.4  &       & 42.0  &       & 31.6  &       & 26.7  &       & 34.1  \\
Customer27-7 &       & 3243.10  &       & 22.1  &       & 20.0  &       & 19.2  &       & 24.5  &       & 24.4  \\
Customer27-8 &       & 3757.85  &       & 124.9  &       & 110.4  &       & 115.4  &       & 92.2  &       & 80.8  \\
Customer27-9 &       & 3607.30  &       & 461.9  &       & 291.4  &       & 214.5  &       & 265.6  &       & 214.4  \\
Customer36-0 &       & 3869.45  &       & 423.3  &       & 386.4  &       & 325.0  &       & 156.2  &       & 175.9  \\
Customer36-1 &       & 4267.15  &       & 407.4  &       & 259.4  &       & 228.9  &       & 56.1  &       & 151.8  \\
Customer36-2 &       & 4543.25  &       & 104.9  &       & 81.2  &       & 81.6  &       & 15.8  &       & 49.4  \\
Customer36-3 &       & 4008.45  &       & 497.8  &       & 283.0  &       & 287.9  &       & 115.8  &       & 293.9  \\
Customer36-4 &       & 4122.55  &       & 1342.3  &       & 1168.0  &       & 814.1  &       & 1342.3  &       & 528.7  \\
Customer36-5 &       & 4511.45 &       & 1245.5  &       & 685.1  &       & 994.7  &       & 868.5  &       & 768.6  \\
Customer36-6 &       & 3913.90  &       & 1505.8  &       & 728.9  &       & 478.1  &       & 1505.8  &       & 304.4  \\
Customer36-7 &       & 4059.90  &       & 112.6  &       & 51.5  &       & 53.0  &       & 50.1  &       & 27.2  \\
Customer36-8 &       & 4722.45  &       & 509.2  &       & 324.0  &       & 393.0  &       & 339.9  &       & 316.3  \\
Customer36-9 &       & 4382.05  &       & 157.4  &       & 98.3  &       & 86.0  &       & 65.9  &       & 60.3  \\
Customer45-0 &       & 5226.25  &       & 2260.5  &       & 1512.1  &       & 1284.0  &       & 1254.4  &       & 1056.8  \\
Customer45-1 &       & 5471.45  &       & 1102.2  &       & 666.5  &       & 841.9  &       & 601.1  &       & 538.9  \\
Customer45-2 &       & 6205.10  &       & 1894.9  &       & 1335.6  &       & 1314.5  &       & 841.4  &       & 783.5  \\
Customer45-3 &       & 5663.10  &       & 1274.5  &       & 813.8  &       & 756.8  &       & 456.8  &       & 539.5  \\
Customer45-4 &  & 5068.45 &  & $\dagger$(36.74\%)  &  & $\dagger$(34.71\%) &  & $\dagger$(0.01\%) &  & 2000.3 &  & 1354.4  \\
Customer45-5 &  & 4935.55 &  & $\dagger$(0.27\%)  &  & $\dagger$(0.10\%) &  & $\dagger$(0.04\%) &  & 2075.9 &  & 1352.9  \\
Customer45-6 &  & 5088.65 &  & 2729.9  &       & 949.3  &       & 1087.6  &       & 1198.8  &       & 1238.5  \\
Customer45-7 &  & 5863.45 &  & 2320.0  &       & 2415.8  &       & 1659.4  &       & 2239.0  &       & 998.4  \\
Customer45-8 &  & 5598.85 &  & $\dagger$(0.14\%)  &  & 2977.2  &       & 2521.6  &       & 2834.1  &       & 2655.7  \\
Customer45-9 &  & 6599.85 &  & $\dagger$(0.17\%)  &  & $\dagger$(0.01\%)  &  & 3549.2  & & 2693.5  &       & 3012.8  \\
    \bottomrule
    \end{tabular}
	}
	\label{ecTable}
\end{table}

Table \ref{ecTable} reports the additional details regarding the computational experiments for the \prob. The column ``Instance'' lists the names of each instance, determined by their problem size and an index ranging from 0 to 9. The following column ``UB'' presents the corresponding optimal value. 
We report the computational times of different configurations of the BPBC algorithm for brevity, where:
\begin{itemize}
	\item Basic: The algorithm described in Section \ref{secAlgorithm} without any enhancement;
	\item Basic-W: Basic enhanced with the warm-start strategies described in Section \ref{secWS};
	\item Basic-WB: BPBC-W enhanced with the bounding-and-fixing strategy described in Section \ref{secBF};
	\item Basic-WBI: BPBC-WB enhanced with valid inequalities (\ref{lowerbound}) described in Section \ref{secLB};
	\item BPBC (Basic-WBIH): BPBC has all enhancements described in Section \ref{secImprovement}, enhancing BPBC-WBI with the primal heuristics in Section~\ref{secUB}.
\end{itemize}
When the BPBC algorithm cannot prove an instance to optimality within the time limit, we mark ``$\dagger$(\%Gap)'' in the corresponding cell with its solution gap.

\end{document}